%% file: ms.tex
\documentclass[11pt,preprint]{aastex}
\usepackage{epsfig,lscape}

\def\spose#1{\hbox to 0pt{#1\hss}}
\def\lax{$\mathrel{\spose{\lower 3pt\hbox{$\mathchar"218$}}
     \raise 2.0pt\hbox{$\mathchar"13C$}}$}
\def\gax{$\mathrel{\spose{\lower 3pt\hbox{$\mathchar"218$}}
     \raise 2.0pt\hbox{$\mathchar"13E$}}$}

\newcommand{\chandra}{{\it Chandra}}
\newcommand{\asca}{{\it ASCA}}
\newcommand{\rosat}{{\it ROSAT}}

\newcommand{\xmm}{{\it XMM-Newton}}
\newcommand{\einstein}{{\it Einstein}}
\newcommand{\lum}{\thinspace\hbox{$\hbox{erg}\thinspace\hbox{s}^{-1}$}}
\newcommand{\flux}{\thinspace\hbox{$\hbox{erg}\thinspace\hbox{cm}^{-2}\thinspace\hbox{s}^{-1}$}}
\begin{document}

\def\spose#1{\hbox to 0pt{#1\hss}}
\def\laeq{\mathrel{\spose{\lower 3pt\hbox{$\mathchar"218$}}
     \raise 2.0pt\hbox{$\mathchar"13C$}}}
\def\gaeq{\mathrel{\spose{\lower 3pt\hbox{$\mathchar"218$}}
     \raise 2.0pt\hbox{$\mathchar"13E$}}}

\title{X-ray Point Sources in The Central Region of M31 as seen by \chandra}
\author{Albert K.H.~Kong, Michael R.~Garcia, Francis A.~Primini,
Stephen S.~Murray, Rosanne Di\,Stefano\altaffilmark{1} and Jeffrey E. McClintock}
\affil{Harvard-Smithsonian Center for Astrophysics, 60
Garden Street, Cambridge, MA 02138}

\altaffiltext{1}{also Department of Physics and Astronomy, Tufts
University, Medford, MA 02155}

\begin{abstract}

We report on \chandra\ observations of the central region of M31. By
combining eight \chandra\ ACIS-I observations taken between 1999 and
2001, we have identified 204 X-ray sources within the central $\sim
17'\times17'$ region of M31, with a detection limit of $\sim
2\times10^{35}$\lum. Of these 204 sources, 22 are identified with
globular clusters, 2 with supernova remnants, 9 with planetary nebula,
and 9 as supersoft sources. By comparing individual images, about
50\% of the sources are variable on time scales of months. We also
found 13 transients, with light curves showing a variety of shapes.
We also extracted the energy spectra of the 20 brightest sources; they
can be well fit by a single power-law with a mean photon index of
1.8. The spectral shapes of 12 sources are shown to be variable,
suggesting that they went through state changes. The luminosity
function of all the point sources is consistent with previous
observations (a broken power-law with a luminosity break at
$1.7\times10^{37}$\lum). However, when the X-ray sources in different
regions are considered separately, different luminosity functions are
obtained.  This indicates that the star-formation history might be
different in different regions.

\end{abstract}


\section{Introduction}
       
At a distance of $\sim 800$ kpc, M31 provides us with a prime
opportunity to study the global properties of a galaxy which is
similar to our Galaxy. As M31 and our Galaxy share similar
morphology and size, comparisons between their X-ray properties can be
enlightening.  Even though M31 is further away, studies of X-ray
sources within it are simplified relative to our Galaxy for several
reasons.  Firstly, the distance is well known, allowing us to
determine X-ray luminosities accurately. Secondly, the extinction of
X-ray sources within M31 is much lower than that typical of sources in
our Galaxy, which enables us to study the X-ray properties of M31 over
a wider energy band. Finally, the moderate inclination angle of M31
allows us to more easily determine the location of X-ray sources
within (or outside of) spiral arms, the galactic bulge, or halo.

M31 has been observed extensively by several earlier X-ray missions.
The first detailed observations were made with the \einstein\
IPC and HRI (Trinchieri \& Fabbiano 1991) and over 100 sources were
detected at a detection limit of $\sim 10^{36}$ \lum\ (0.2--4.0
keV). Subsequent \rosat\ HRI observations of the central $\sim 34'$
revealed 86 sources at a similar detection limit (Primini, Forman, \&
Jones 1993, hereafter PFJ93). A comparison of the \einstein\ and
\rosat\ source lists showed that $\sim$42\% of the sources within the
central $7.5'$ region were variable.  Recently, two deep and extensive
\rosat\ PSPC surveys have identified 560 point sources in the entire disk
of M31 (Supper et al. 1997, 2001; hereafter Su97, Su01). The detection
limit of this survey was $\sim 5\times 10^{35}$ \lum\ (0.1--2 keV).  
These 560 sources include 33 globular clusters, 16 supernova
remnants, 15 supersoft sources, 55 foreground stars and 10 background
objects.

M31 was observed by \chandra\ and \xmm\ soon after these observatories
were launched. In
the first observation (8.8 ks) of the core region by \chandra\ in 1999
October, 121 point sources were identified within the central
$17'\times17'$ region and the nucleus, which had been seen as one source by
previous missions, was nicely resolved into five point sources 
(Garcia et al. 2000a). Moreover, a bright transient was
discovered $\sim 26''$ from the nucleus. A relatively deeper \xmm\
observation (34.8 ks) was made in 2000 June; 116 sources were detected
down to a limiting luminosity of $6\times 10^{35}$ \lum\ (0.3--12 kpc;
Shirey et al. 2001) and a pulsating supersoft transient with a
periodicity of $\sim 865$ s was discovered (Osborne et
al. 2001). Moreover, both \chandra\ (Garcia et al. 2001a; Primini et
al. 2000) and \xmm\ (Shirey et al. 2001) observations confirmed that
the unresolved X-ray emission in the core region is much softer than
most of the resolved X-ray sources in that region. Fifteen X-ray
point sources are newly associated with globular clusters (Di\,Stefano
et al. 2002), which when combined with the previous \rosat\ results
(Su01) brings the total number of M31 globular clusters with
detected X-ray emission to 48.  The X-ray luminosity of these globular
clusters appears to be substantially higher than clusters in our Galaxy;
for example, in M31 $\sim 1/3$ of the clusters have luminosities $L_X >
10^{37}$ \lum\ (0.5--7.0 keV) as compared to $\sim 1/12$ within the
Galaxy (Di~Stefano et~al. 2002).  \chandra\ and \xmm\ also discovered
several bright ($L_X > 10^{37}$ \lum) transients in M31 and M32
(Garcia et al. 2000a; Garcia et al. 2000b; Osborne et al. 2001; Shirey
2001; Kong et al. 2001; Garcia et al. 2001b).  The brightest of these reached a peak
luminosity of $L_X \sim 3\times10^{38}$ \lum.

We report herein the properties of the point sources in the central
$\sim 17'\times 17'$ region of M31 as deduced from eight separate
$\sim 5$ks ACIS-I observations spanning $\sim 1.5$~years.  
The detection limit (while variable) is  $\sim 2 \times
10^{35}$erg~s$^{-1}$ (0.3--7 keV) across most of this region.  In addition
to a source catalog, we discuss the overall spectral properties,
temporal variability and luminosity function of these sources.  The
brighter sources have sufficient counts to allow meaningful searches
for spectral variability, allowing comparison to Galactic
sources. The main focus of this paper is on the complete source list
and the overall properties of point sources in M31.  In a series of
companion papers, we will discuss the diffuse emission (for instance,
see Primini et al. 2000 and Dosaj et al. 2001 for a preliminary
analysis), the X-ray emission from the central supermassive black hole
(Garcia et al. 2002), temporal and spectral variability (Kong et al., in
preparation) and supersoft source populations (Di\,Stefano et al., in preparation)

In this paper, we adopt a distance of 780 kpc (Stanek \& Garnavich
1998; Macri et al. 2001) and assume a hydrogen column density 
equal to the Galactic value of $\sim 7\times10^{20}$ cm$^{-2}$ (Dickey
\& Lockman 1990) unless otherwise specified. The quoted errors
throughout this paper are at 1-$\sigma$ confidence, unless otherwise specified.

\section{Observations and Data Reduction}

M31 was observed with \chandra\ regularly as part of the AO-1 and AO-2
GTO program during 1999--2001. The program was designed to search for
transients.  The observations consist of a series of HRC snapshots
($\sim$ 1 ks) that cover the entire galaxy.  If a transient is
discovered in the HRC mosaic, then a follow-up ACIS image ($\sim$ 5
ks) of the transient is obtained; otherwise an ACIS image of the
nucleus is obtained.  In this paper we focus only on the ACIS-I (I0,
I1, I2 and I3) data obtained for the central $16.9' \times 16.9'$
region of M31.  These data consist of 8 separate observations, with
exposure times ranging from 4 to 8.8 ks. The details of the
observations are given in Table 1. The nuclear region of M31 was
placed near the aim-point of the ACIS-I array. The focal-plane
temperature was $-110^{\circ}$C during the first four observations,
and $-120^{\circ}$C for the others. The observations were made at
various spacecraft roll angles; consequently, the total region covered
by the observations is slightly larger than $16.9' \times 16.9'$, and
sources near the outer edge of ACIS are not observable in all eight
exposures.

All data were telemetered in Faint mode and were collected with a frame
transfer time of 3.2 s. In order to reduce the instrumental
background, only data with \asca\ grades of 0, 2, 3, 4, and 6 were
included. We selected data free from bad columns, hot pixels and
columns close to the borders of each ACIS chip node. The standard
$0.5''$ pixel randomization was also removed. Only events with photon
energies in the range of 0.3--7.0 keV were included in our analysis. We
inspected the background count rates from the ACIS-S3 chip for all of
the observations; except for the first observation, no significant
background flares were found. For the first observation, periods with
high background were rejected, resulting in an effective exposure time of
8.8 ks out of 17.5 ks (see Garcia et al. 2000a for details of this
observation).

The data reduction and analysis was done with CIAO v2.1
\footnote{http://asc.harvard.edu/ciao/}. Some of the image analysis
was done with IRAF \footnote{http://iraf.noao.edu/}, while the spectra
were analyzed with XSPEC v11
\footnote{http://heasarc.gsfc.nasa.gov/docs/xanadu/xspec/index.html}. 

\section{Analysis}

\subsection{X-ray Images}

In order to create a deep image suitable for the detection of faint
sources, the eight observations were combined into a single stacked
image.  The rms uncertainty of the aspect solution of \chandra\ is
$\sim 1''$, but there can be systematic errors (up to $2''$) for
specific ACIS-I observations \footnote{see
http://asc.harvard.edu/mta/ASPECT/}.  These errors are larger than the
PSF, and therefore could degrade the PSF of the stacked image.  We
removed these errors by registering all the data sets to the
coordinate frame of OBSID00312 using 12 bright X-ray sources within
$\sim 5'$ of the aim-point.  The positions of these sources were
determined from the weighted centroid of the counts in a $0.5''$ pixel
(= one ACIS pixel) image.  The cross registration of the eight data
sets is accurate to $\sim 0.4''$ (as measured from the rms differences
in the shifts of the 12 sources). 

Because the nuclear region of M31 is particularly crowded (see Garcia
et al. 2000a), we were concerned that the $\sim 0.4''$ accuracy
achieved above would be insufficient.  We therefore repeated the
procedure above on an image of the central $2'\times2'$, but used a
pixel size of 
$1/8''$ (= 1/4 ACIS pixel) scale.  
Such sub-pixel images can easily be generated using CIAO routines.  Any
\chandra\ image in sky co-ordinates must be reconstructed from the
instantaneous space-craft dither position, and the arrival times and detector
locations of the individual photons.  While reconstruction of the image
on a scale that matches the detector pixel size is a natural choice,
one can just as easily reconstruct the image a finer scale.  Of course,
the larger size of the detector pixel means that the photon location
within this finer scale is randomized by the dither position at the
photon arrival time.  When we generated a set of sub-pixel images from
our 8 observations, we found that we were able to reduce the rms
registration errors in the stacked image to $0.25''$.  
 
The {\it absolute} astrometry of these images is still limited by the
\chandra\ aspect solution to $\sim 1''$.  We choose to stack the
images in the coordinate frame of OBSID00312 merely because it was
convenient.  The systematic error of this coordinate frame (processing
system version R4CU5UPD8.2) should be $\sim 1''$.  While future
reprocessing of all \chandra\ data to remove systematic aspect errors is
expected to yield absolute astrometry of $\sim 0.5''$,  more accurate
astrometry may be possible by registering the \chandra\ image with astrometric
catalogs.  This has been done by Kaaret (2002) with an HRC image of M31
and the 2MASS catalog to an accuracy of $<0.4''$ (see Kaaret 2002 for
details).  The positions listed in Table~2 and throughout this paper 
use this astrometric
reference. 

We note that the exposure of the combined image ranges from 
4~ks (at the extreme edges) to 39.7~ks (in the center). More than 170
sources have a total integration time of 30~ks.

Figure 1 shows the stacked ``true color'' image of M31, which is a
composite of images from soft (0.3--1.0 keV), middle (1--2 keV) and hard
(2--7 keV) bands. Soft sources appear red, moderately hard sources
appear green, and the hardest sources appear blue. The image has been
corrected for exposure and smoothed slightly with a  Gaussian ($\sigma=1.96''$)
in order to improve the appearance of point sources.  In Figure 2, we
show the ``true color'' image of the central $2'\times2'$ region of M31
in $1/8''$ pixel resolution, with the possible nuclear counterpart
(M31$^*$) marked (Garcia et al. 2002).

\subsection{Source Detection}

Discrete sources in the stacked image were found with WAVDETECT
(Freeman et al. 2002), a wavelet detection algorithm implemented
within CIAO. The central $2'\times2'$ region was treated separately by
using the $1/8''$ pixel image. The two source lists were combined into
the master source list shown in Table 2. We set the detection
threshold to be $10^{-6}$ and carried out the wavelet analysis at 9
scales (1, $\sqrt{2}$, 2, $2\sqrt{2}$, 4, $4\sqrt{2}$, 8, $8\sqrt{2}$
and 16). This detection threshold corresponds to $< 1$ false detection
due to statistical fluctuations in the background. 
We then repeated this procedure on the annular square region between $2'$
and $8'$ in size using an image with $1/2''$ pixels. On the region
outside of the $8'\times 8'$ square we used an image with $2''$ pixels. 
A total of 205
sources were detected.  Source count rates were determined via
aperture photometry.  The radius of the aperture was varied with
average off axis angle in order to match the 90\% encircled energy
function.  The average off axis angle was computed for each
source based on the off axis angles of the eight observations weighted
by their exposure times. The extraction radius varies from $\sim 1''$ near
the aim-point to $\sim 17''$ for the sources with the largest off axis
angles.

Background was extracted from an annulus centered on each source. In
some cases, for example in the nuclear region, we modified the
extraction region to avoid nearby sources. It was also necessary to
modify the extraction radius for some faint sources that are close to more
luminous sources. Every extraction region was examined carefully in
the image. The count rate was corrected for exposure, background
variation and instrumental PSF. We examined the image of each
detected source, and found only one case in which the detection 
was clearly spurious (this source had signal-to-noise ratio S/N $< 2$).  

Table 2 lists the 204 sources in our catalog, sorted in order of
increasing R.A.  The columns give the
source number, the IAU approved name, the position (J2000), the net
counts, the count rate and 1-$\sigma$ error, hardness ratios (see
\S3.4) and the 0.3--7.0 keV luminosity.  The source numbers have a
prefix of r1, r2 and r3.  The r1 sources are located in the central
$2'\times 2'$ or ``inner bulge'' region. The r2 sources are located
within the central $8'\times8'$ excluding the inner bulge region; we
refer to this annular region as the ``outer bulge''. The r3 sources
are located outside the central $8'\times 8'$ in the ``disk'' region.
This nomenclature  is based on optical studies (Morton,
Andereck \& Bernard 1977); a similar classification was also used by
Trinchieri \& Fabbiano (1991). The conversion to luminosities assumes
an absorbed power-law spectrum with a photon index of 1.7 and
$N_H=10^{21}$ cm$^{-2}$, which is the typical spectrum of a point
source in M31 (Shirey et al. 2001).  All sources in the catalog have
S/N $> 2.5$ and only 5 have S/N $< 3$.  The detection limit for the
sources varies across the image due to the variations of exposure
time, background and instrumental PSF, and is highest near the edges
where the PSF broadens rapidly and exposure time is lowest.  Over the
inner $4'$ of the field, the detection limit is $2.1\times 10^{-4}$
counts s$^{-1}$, which is equivalent to $L_X \sim 2\times 10^{35}$ erg
s$^{-1}$.

\subsection{Source Identification}

Tables 3--5 summarize the results of matching our new ACIS source
catalog with existing catalogs of M31 objects.
Table 3 lists the number of matches found in each of the catalogs we
searched, and also the number of accidental (spurious) matches expected.
Table 4 lists the catalog identifications of the ACIS
sources and the radial offset between the cataloged object and the ACIS
source. Table 5 itemizes the
matches between our ACIS catalog and the \rosat\ HRI source list
(PFJ93).   We varied the search radius based on the accuracy of the
various catalogs (increasing it as needed), and also based on the density of
sources in the catalogs (reducing it in order to limit accidental
matches). 

We used the following catalogs and corresponding search radii:

{\it ROSAT sources}: the \rosat\ HRI catalog (PFJ93) --- $6''$
search radius.

{\it Globular clusters}: the Bologna catalog (Battistini et al. 1987),
the catalog by Magnier (1993), and a recent catalog based on
{\it HST} data by Barmby (2001) ---   $3''$ search radius for all
three. 

{\it Supernova remnants}: the lists by d'Odorico et al. (1980), Braun
\& Walterbos (1993), and Magnier et al. (1995) --- $10''$
search radius.

{\it Planetary nebulae}: the catalogs by Ford \& Jacoby (1978) and
Ciardullo et al. (1989)---  $3''$ search radius.

{\it Extragalactic objects}: the NASA/IPAC Extragalactic Database
(NED) and SIMBAD --- $3''$ search radius.

{\it M31 stars}: the catalog of 485,425 objects (mainly stars) by Haiman et
al. (1994) and the SIMBAD --- $0.8''$ search radius.

{\it Stellar Nova}: as reported in the IAUC during the period covered
by our observations --- $3''$ search radius.

{\it OB Associations}: the catalog of 174 OB associations within M31
identified by Magnier et al. 1993) --- $3''$ search
radius. 

In order to estimate the accidental correlation rate, we used a
technique similar to that described by Hornschemeier et al. (2001). We shifted
all the \chandra\ sources by $10''$ to the northeast, northwest,
southeast, and southwest, and ran the search for each of the
catalogs. The results were averaged to estimate the accidental
matching rate listed in Table 3.  With the exception of the large
catalog of M31 stars, the accidental matching 
rates are generally small, therefore justifying our choice of search 
radius.  For example, the average accidental matching rate
with \rosat\ sources is 9.5, which is about $10\%$ of the total number of
\rosat\ matches found (77).  

We find two matches with SNR, but predict a random match rate (1.0)
that is similar to what we find.  However, the unusual nature of the
two sources which match with the SNR catalog leads us to believe that
these matches are real.  Both of these \chandra\ sources have
relatively low hardness ratio values (see \S3.4) and \chandra\ source
r3-63 is resolved into a ring-like object (Kong et al. 2002).  Both of these
identifications are listed in Table~4, and the total number of true
SNR IDs is listed in Table~3 as two.

We found 29 \chandra\ sources (excluding optically identified globular
clusters and planetary nebulae) within $0.8''$ of objects in the Haiman et
al. (1994) catalog, which consists mainly of stars in the field of
M31.  The number of accidental matches even with this small $0.8''$
search radius is 21.25, nearly equal to the number found!  This is due
to the very high density of objects in this catalog, and leads us to
expect that most of the 29~matches are spurious.  We note that our
$0.8''$ searching radius is roughly equal to the astrometric accuracy of the Haiman et
al. catalog.

However, one can test these
possible identifications with the spectral (or hardness ratio) data.
X-ray emission from stars is relatively soft; for instance, the PSPC
survey (Su97) showed that the energy spectrum of a foreground star can
be best fitted by a Raymond-Smith model with $kT_{RS}\sim1$ keV or a
power-law model with a photon index $\alpha \gaeq 3$. However the
\chandra\ sources which have a sufficient number of counts for
spectral analysis (see \S3.5) have much harder spectra ($\alpha \sim
1-2$) and are therefore unlikely to be stars.
In the remaining cases where there are too few
counts we examined the hardness ratio (HR2; see \S3.5). If the ratio
is inconsistent with a soft spectrum (i.e., if HR2 is greater than
$-0.5$), we similarly rule out a stellar association.  In this way we
rule out 25 of the possible 29 matches.  Table~4 lists the remaining 4
possible stellar identifications.  We note the (fortuitous?)
agreement in numbers between the number of sources with soft spectra
(4) and the accidental matching rate which leads us to expect that
$\sim 7$ out of our possible 29~matches may be real.  These numbers
are summarized in Table~3.

We find that 77 \chandra\ sources have counterparts in the \rosat\ HRI
catalog (see Table~5).  The remaining 127 ($=204-77$) \chandra\ sources
were not detected in the \rosat\ HRI catalog, presumably because they
are below the \rosat\ detection limit or are variable.  The \chandra\
catalog extends $\sim 5\times$ fainter than the \rosat\ HRI catalog.
We have not compared our \chandra\ catalog to the new complete \rosat\
PSPC catalog (Su01) because the later catalog covers a much larger
area and has substantially lower resolution, making it less appropriate
to compare to this ACIS data than the HRI catalog.  We note that in 6
cases
the \rosat\ HRI sources match up with 2 or more \chandra\ sources,
indicating either that these \rosat\ sources have been resolved by \chandra\, or
that one (or more) of the \chandra\ sources is transient. It is worth
noting that PF93(44) is the nucleus identified by PFJ93.
Alternatively, we note that we expect $\sim 9.5$ spurious matches with
\rosat\ sources, so something like $\sim 10$\% of these multiple
matches may be spurious.  The expected numbers of spurious matches
leads us to list the ``true'' number of matches as 67 in Table~3.

We identify 22 \chandra\ sources with globular clusters (see Table~4).
Eight of these globular clusters are identified as X-ray sources
(comparing to previous \rosat\ observations) for
the first time, while three of them are new to the recent studies by
\chandra\ (Di\,Stefano et al. 2002) because of the different roll
angles and improved astrometry. Because globular clusters are sometimes used to
register M31 images taken at different wavelengths, it may be worth
noting that the average radial offset (after taking the average
offsets in RA and DEC) between the X-ray and optical positions is
$0.77''$.
\chandra\ source r3-74 is within $1''$ of both mita165 and
mita166, but we note that these clusters themselves are separated by
only $\sim 1.5''$.  We therefore allow the possibility that in reality
they constitute a single cluster.  This possibility, along with the
number of expected spurious matches, leads us to list the number of
detected globular clusters as 22 in Table~3.

We found 11 matches between \chandra\ sources and planetary nebula
(PN), and expect that $\sim 5$ of these are spurious matches. 
Ford 316 is within $2.4''$ of both r1-21 and r1-33, both of which are
in the very crowded nuclear region of M31.  It seems likely that
Ford 316 is associated with only one (but not both) of these \chandra\ sources. 
\chandra\ source r1-26 is within $3''$ of both Ford 21 and Ford 74,
but it  seems unlikely that both PN would be associated with a single
X-ray source. While it is unclear which associations might be
spurious, these multiple identifications and the expected number of
spurious sources leads us to estimate that there are 9~true
associations between PN and \chandra\ sources (see Tables 3 and 4).

Planetary nebula in our Galaxy are in general rather weak and soft
X-ray sources, with $L_X \sim 10^{30}$erg~s$^{-1}$ and kT$\sim
0.5$~keV (Guerrero et al. 2000, 2001).  The most luminous PN within our
galaxy has $L_X \sim 1.3 \times 10^{32}$erg~s$^{-1}$ and kT$\sim
0.3$~keV (Kastner, Vrtilek, \& Soker 2001).  These \chandra\ sources have
luminosities ranging from $2 \times 10^{35}$erg~s$^{-1}$ to $5 \times
10^{37}$erg~s$^{-1}$ and (with the exception of r2-56) X-ray colors
much harder than Galactic PN, raising the possibility that these
sources are either very unusual or something other than PN.  We note that
the Ford \& Jacoby (1978) survey identified PN on the basis of on and off band
imaging within the 5007\AA~ O [III] line.  Symbiotic stars often show
strong 5007\AA\/ emission and are generally soft X-ray sources of
modest intensity, with the single exception of GX 1+4.  This object
has a hard spectrum and $L_X \sim 10^{37}$erg~s$^{-1}$ because of its neutron star (as
opposed to white dwarf or main sequence) primary.
This suggests that these ``PN'' may
in fact be GX 1+4 analogs in M31.  Optical spectroscopy could help
determine the nature of these sources.

We searched for matches between stellar novae as listed in the IAUC
that were
contemporaneous with our \chandra\ observations we
found no matches within $3''$.  Because novae may appear as X-ray
sources some time after their optical appearance, we started our search with the
nova reported in IAUC
7093 (Stagni, Buonomo, \& di Mille 1999 Jan), which appeared $\sim 10$ months before our first
\chandra\ observation.  Other novae include those in IAUC 7218 (Modjaz
and Li 1999 July), 7236 (Johnson, Modjaz, \& Li 1999 Aug), 7272 (Filippenko et
al. 1999 Oct), 7477 (Li, 2000 Aug), 7516 (Donato et al. 2000 Nov) and
7709 (Fiaschi, Di Mille, \& Cariolato 2001 Sept).
While outside of our search radius ($3''$), a nova found in 2001 September
(IAUC 7709, Fiaschi, Di Mille \& Cariolato 2001) is within $5''$ of
source r2-29.  This \chandra\ source is positionally coincident with
XMMU J004234.1+411808, which Osborne et al. (2001) suggested may be
an X-ray nova.  This source was not detected in the ACIS observation
of 2000 June 1 (OBSID 309); it reached a peak flux in an HRC-I observation on
2000 June~6 (OBSID 273),  more than one year before the optical nova
(see Figure 5).  
We know of no cases in which the X-ray outburst of an optical nova (or
an X-ray nova) proceeded the optical outburst by $\sim 1$~year, which
therefore suggests that the spatial proximity of the two events is a
random coincidence. The expectation value of such events is 0.25, so
this is moderately possible. 

We also searched for matches with OB associations, as O and B stars
may be moderate X-ray sources ($L_X < 10^{33}$erg~s$^{-1}$, Berghofer
et al. 1997).  While this is well below our detection limit, a group
of O and/or B stars may reach our detection threshold,  and star
forming regions could conceivably harbor massive X-ray binaries.
However, we found no matches within our search radius of $3''$.

We searched for extragalactic emission line objects as listed in NED
and SIMBAD, and found a single match (r3-83, see Table 4).  AGN are
likely to be obscured by the bulge of M31 and therefore we expect that
we may detect many AGN without optical counterparts.  A better
estimate to the numbers of such objects is made based on deep field
observations in the next section.

\subsection{Hardness Ratios}

Many of the sources in our catalog have \lax~100 counts, which makes
it difficult to derive spectral parameters with meaningful
constraints.  However, hardness ratios can give a crude indication of
the X-ray spectra in these cases.  We therefore computed
the hardness ratios for all the detected sources.  These ratios were
based on the source counts in three energy bands: S (0.3--1.0 keV), M
(1--2 keV) and H (2--7 keV). The two hardness ratios are defined as
HR1=(M-S)/(M+S) and HR2=(H-S)/(H+S). We calculated the uncertainties
of the hardness ratios by using maximum likelihood method as used in
the {\it Einstein} catalog (Harris et al. 1993).
Table 2 lists both HR1 and
HR2. Figures~3 and 4, respectively, show the color-color diagram (CD) and
hardness-intensity diagram (HID) for sources with over 20 counts.
We have overlaid the CD with 6~lines showing the tracks followed by
representative  spectra with differing values of ${\rm N_H}$.
Power-law spectra tend to occupy the top right section of the diagram,
while soft blackbody models occupy the lower left. 
For example, a
``supersoft source'' (SSS) having a blackbody spectrum with a
temperature of 70 eV and $N_H$ of $10^{21}$ cm$^{-2}$ would be in the
lower left with HR1=$-0.98$ and HR2=$-1$.
A typical AGN or X-ray binary with a power-law spectrum and a photon index 
$\alpha=1.7$ would be in the extreme upper right if ${\rm N_H} \geq
10^{22}$ cm$^{-2}$.

From the CD in Figure~3, it is clear that there are three sources with
more than 20~counts and with both HR1 and HR2 consistent with $-1$
(r3-84, r2-12, and r3-8).  This indicates that they have no counts above
1~keV, and therefore are very likely SSSs.  
The brightest of these three sources (r2-12) is within the central
$8'\times8'$ region and has $L_X >10^{37}$ erg s$^{-1}$. This source
was also detected in the \rosat\ surveys and identified as a SSS
(Su01).  

We searched our catalog for other candidate SSSs, following a method
analogous to that of Su97 and Kahabka (1999).  These surveys
identified 15 SSSs (Su97) or perhaps 16 additional SSSs (Kuhabka 1999), depending
upon the selection criteria used.
Candidate \chandra\  SSSs are
those with HR2+$\sigma_{HR2} \leq -1$ {\it and} HR1 $< 0$, {\bf or}
HR1+$\sigma_{HR1} \leq -0.8$.  There are 14 sources satisfying these
conditions: r1-25, r2-12, r2-19, r2-42, r2-46, r2-56, r2-57, r3-5,
r3-8, r3-63, r3-69, r3-77, r3-84, and r3-96.  A single one
of these (r1-25) is in the central $2'\times2'$ region. Source r2-12 was
noted as a bright SSS in both the {\it Einstein} and PSPC surveys.  Source
r3-8 was not classified as a SSS in the PSPC surveys (Su97, Su01) but
was subsequently identified as a SSS based on the less
tight selection criteria of Kahabka (1999).  Sources r3-63 and r3-69
were both detected in the PSPC survey and were identified to be SNRs
(see Table~4).  Source r2-42, r2-46 and r2-19 may be associated with 
stars listed in the Ha94 catalog.  Removing these stars and the two SNRs
from consideration, we are left with 9 SSS candidates, 7 of which
are newly discovered.

In the CD and HID, we have marked the three regions of M31 we have
defined in different colors:
the central $2'\times2'$ (region~1) is blue, the central
$8'\times8'$ excluding the central $2'\times2'$ (region~2) is red, and
the entire field excluding the central $8'\times8'$ (region~3) is green.
All three regions show extensive overlap in both diagrams, with one
clear exception: the top right hand corner of the CD is populated
mainly 
with sources from region~3.  The appearance of this figure is born out
by the average hardness ratios for the three regions as listed in
Table~6. Both HR1 and HR2 of region~3 are significantly ($\sim
5\sigma$) higher than the values for regions~1 and 2, which are
themselves consistent at the $\sim 2\sigma$ level.  Although these
sources may be intrinsically hard and within M31, many are probably
strongly absorbed background AGN.  

These harder sources may form a distinct group in the CD, i.e., there
may be a separate clump of sources with HR1~$>0.6$ and HR2~$>0.5$.
There are 48 sources that meet these hardness ratio criteria, 42 of
them are in region~3.  Based on the \chandra\ deep field observations
(Brandt et al. 2001), we estimate that there should be $\sim 30$
serendipitous sources in our field.  Thus, the majority of these will be
background AGN and would therefore be strongly absorbed by the dust
and gas in M31.

\subsection{Temporal Variability}

The eight \chandra\ ACIS-I observations described herein span nearly 2
years from 1999--2001. This is substantially longer than previous
surveys by \rosat\ (2 observations separated by $\sim 1$ year).  In
order to study long-term X-ray variability, we computed a variability
parameter following PFJ93:

\begin{eqnarray}
S(F_{max}-F_{min})=\frac{|F_{max}-F_{min}|}{\sqrt{\sigma^{2}_{F_{max}}+\sigma^{2}_{F_{min}}}},
\end{eqnarray}

\noindent where $F_{min}$ and $F_{max}$ are the minimum and maximum X-ray flux
during the 2 years of observations, and $\sigma_{min}$ and $\sigma_{max}$
are the corresponding errors. We define a source to be a variable if
$S > 3\sigma$.  The 100 sources we found to be variable using this
criteria are indicated with a ``v'' in Table~2. We note that 6 of the
sources are excluded from this analysis as they were only observed
once.  Thus, the variable sources represent $\sim50$\% of the total.
The minimum amount of variability found corresponds to a factor of 1.5.

We note that the fraction of variable sources depends upon the region
of M31 considered.  In the central $2'\times2'$ region, 73\% of the
sources are variable, while this fraction drops to 58\% and 39\% in
regions~2 and~3, respectively (see Table~7 and Figure 5).  One might worry that
this apparent trend is an artifact of the variation in PSF and/or
diffuse background with off axis angle.  The average number of
background counts per source in regions 1 and 2 is $<3$, while in
region 3 the average is 12. These background counts could mask small
variations in source flux.  In order to
determine the effect of the variable PSF, we used
MARX\,\footnote{http://space.mit.edu/ASC/MARX/} to simulate this
effect.  The maximum and minimum intensity images of the 24~variable
sources found in region~1 were re-projected to random positions within
region~3.  The observed background (with Poisson noise) was
included. We then re-computed S for these sources, and find that the
average change from the original numbers is $\sim 2\%$.  This
indicates that the measurement of variability is robust.

If a source is faint, the fractional error of its
flux will be larger than that of brighter sources. Therefore,
a larger fractional change in flux is required to satisfy $S >
3\sigma$.  If the fractional change in flux is independent of the
source luminosity, Eqn. 1 will be biased against finding faint sources
with $S > 3\sigma$.  
Given that regions 2 and 3 have a larger proportion of faint sources
than region~1, part of the apparent deficit in variable sources in
regions 2 and 3 may be to this bias. 
This possible bias against faint sources can be removed by simply
excluding faint sources from consideration.  Figure 5 shows the
fraction of X-ray variables in each of the regions, computed for three
progressively higher low luminosity cutoffs (the first being the
detection limit).  Limiting ourselves to sources with $L_X >
10^{36}$\lum, we see that that the fraction of variables in region 3 is
still lower than that in regions 1 and 2, but that region 2 now has
the highest fraction of variables. At the highest cutoff that still
gives a reasonable number of sources ($> 10^{37}$\lum), region 3 still
has a smaller fraction of variables than region 2.  Given that the
luminosity distribution is similar in these two regions (see \S\,4 for
details) the smaller fraction of variables in region 3 appears to be
an intrinsic property of these sources.

Previous observations may be consistent with this apparent trend.  For
example, by comparing \rosat\ observations to \einstein\ observations
made 10~years earlier, PFJ93 found that $\sim 42$\% of the X-ray
sources in the central $7.5'$ region were variable.  By comparing two
\xmm\ observations separated by six months Osborne et al. (2001) found
that $>15$\% of the sources in the central $30'$ were variable.

We have also discovered 12 bright transients which
are indicated with a ``t'' in Table~2. We define a bright transient as
follows: i) the source has $S>3\sigma$ {\it and} ii) the source is
found in at least one observation with a luminosity of $ \gaeq
5\times10^{36}$ erg s$^{-1}$ and is not detected (ie, the counts at
the source location are below the $3\sigma$ detection threshold) in 
at least one other observation. 
Note that the
luminosity limit covers typical outburst luminosities of soft X-ray
transients and Be/X-ray binaries in our Galaxy.

One important transient was missed by this analysis because it had a
peak luminosity below our ``bright transient'' threshold during the
eight ACIS-I observations considered here.  This
object is XMMU J004234.1+411808 (= CXOM31 J004234.4+411809 = r2-29),
which Osborne et al. (2001) and Trudolyubov, Borozdin, \& Priedhorsky
(2001) suggest is an X-ray nova.  Observations with the HRC-I (OBSIDs
273, 275, and 276), the ACIS-S (OBSIDs 309 and 310) and \xmm\ (Osborne
et al. 2001) allow us to reconstruct the lightcurve (Figure~6a),
showing that this object had a peak $L_X \sim 2.4\pm 0.5 \times
10^{37}$erg~s$^{-1}$ and an exponential decay with a $\sim 20$~day
e-folding time.  This source increases the number of bright transients
discovered in the last two years within the central region of M31 to 13. 

Two of these transients show typical lightcurves with a fast rise
followed by an exponential decay (r2-29 and r2-28; see Figure 6a,b)
like those seen in our Galaxy (Chen, Shrader \& Livio 1997). A
different sort of behavior is shown by the transient discovered with
\chandra\ in the first observation (r1-5; Garcia et al. 2000a), which
remained in outburst for more than one year and finally turned off in
2001 June (Figure 6c).  We note that Figure~6 includes data from the
HRC-I and ACIS-S which we do not analyze herein beyond measuring
counting rates and luminosities for these few sources. The temporal
variability of the 13 transients and other sources will be described
on a future paper which will fully utilize the HRC and ACIS data (Kong
et al., in preparation).

One source, while not a bright transient, deserves special mention
because of it's unusual long-term variability.  Source r3-44 (Figure~6d)
is in the M31 globular cluster Bo86 and shows a possible $\sim 200$
day modulation.  This is reminiscent of the Galactic source
4U\,1820--30 in the globular cluster NGC~6624, which
has a 176-d long-term modulation (e.g. Bloser et al. 2000).

\subsection{X-ray Spectra and Spectral Variability}

We extracted the energy spectra of the brightest 20 sources from the
first observation (OBSID 303) and fit them to simple
one-component models consisting of absorbed power-law, blackbody and
Raymond-Smith (RS) shapes.   In this paper we limit ourselves to the spectra
as seen in this longest single observation (8.8 ks) because of the
complications involved in fitting the spectrum of a source which
appears at different detector positions and is observed at differing
ACIS temperatures (i.e., with differing detector response matrices).
These brightest 20 sources all have $>300$ detected counts, and are
well distributed across the field.  They range from within $\sim20''$
of the nucleus to over $7'$ distant.
Circular extraction regions centered on the source positions
were applied and correspond to 90\% encircled energy. In order to employ
$\chi^2$ statistics to be used, all spectra were grouped into at least
20~counts per spectral bin. 

All the sources were satisfactorily fit with simple absorbed power-law
or RS models.  Blackbody models gave very poor fits ($\chi^2_{\nu} >
2$) in most cases, except for sources r3-42, r3-52 and r3-61. Table~8
lists the best fitting parameters determined by the power-law and RS
fits to these 20 sources. The power-law photon indices of the 20
sources range from 1 to 3 with a mean of 1.8.  The average and minimum
RS temperatures (kT$\sim 16$~keV and kT$>2.4$~keV, respectively) are
both higher than that of coronal sources (Dempsey et al. 1993),
indicating that none of these 20~sources is likely associated with a
foreground star.

Source r2-26 was the only one of these twenty with a single-pixel
counting rate high enough so that ``pile-up'' could cause an
erroneously hard spectral fit.  A fit to all the data for this source
finds a rather hard spectrum of $\alpha=1$.  In order to mitigate the
effects of pileup we also extracted the spectrum in an annulus around
the heavily piled-up core, and find a softer spectrum of $\alpha=1.6$
(see Table~8).  This slope is in good agreement with a fit to all the
data using a model which attempts to correct for the effects of pileup
(Davis 2001).  In addition, a high S/N spectrum obtained from an
archival \xmm\ /PN data
(taken on 2000 June 25) also
shows similar results, confirming that the correct spectrum of r2-26
is very typical among M31 X-ray sources.

The column density for these 20 sources ranges from effectively zero
to $9\times10^{21}$ cm$^{-2}$ and has an average value of ${\rm N_H} =
2\times 10^{21}$cm$^{-2}$.  The Galactic absorption along the line of
sight to M31 is $\sim 7\times10^{20}$cm$^{-2}$ (Dickey \& Lockman
1990), indicating that most of these bright sources have some
additional local absorption. Figure~7 shows the spectra of the softest
($\alpha=3$) and the hardest ($\alpha=1.1$) of these sources.

We searched for spectral variability in all of the sources using a
method analogous to that described in \S3.5, but replacing $F$ in
equation (1) with HR2.  If this newly defined $S$ is greater than
$3\sigma$, the source is identified as a spectral variable and noted
as ``sv'' in Table~2.

Only 12  sources are found to meet our criteria for spectral
variability, corresponding to 6\% of the total population.  This is of
course a lower limit, as small changes in the spectra of weak sources
are undetectable with the number of counts accumulated in our 40~ks of
merged data.

In order to further investigate the nature of the spectral variations,
we fit simple spectra to two of the brighter spectral variables.  The
fits show that as the counting rate increases, the spectrum becomes
harder (see Figure~8).  This is reminiscent of atoll and Z sources in
our Galaxy (see e.g. Hasinger \& van dar Klis 1989). The luminosity of
these two sources ranges from $(0.4-1.0)\times 10^{38}$ erg s$^{-1}$,
which is higher than the typical luminosity 
of atoll sources ($< 10^{37}$ erg s$^{-1}$).  However, this luminosity
is similar to that of the
Z~sources, which are believed to reach the Eddington limit (Psaltis,
Lamb, \& Miller 1995).  The
luminosity and spectral changes appear consistent with that seen in
Z sources as they move along the ``normal branch''.  
We suggest that these
sources are among the first examples of possible extra-galactic
Z-sources found (along with LMC X-2, Smale \& Kuulkers 2000).
Continued
monitoring may confirm the nature of these sources by revealing the
full Z-shape of the spectral variations observed for galactic Z sources.

\section{Luminosity Function}

The count rates for sources were converted into unabsorbed
0.3--7.0~keV luminosities by assuming an absorbed power-law model with
${\rm N_H}=10^{21}$ cm$^{-2}$ and $\alpha=1.7$.  This is a median spectrum,
as can be seen from Figure~3.  The resulting conversion between count
rate and luminosity is $7.5\times10^{38}$ erg
count$^{-1}$. Luminosities using this conversion are listed in
Table~2.  The derived luminosities are not very sensitive to the
assumed spectral parameters, for example varying ${\rm N_H}$ from
$(5-20)\times10^{20}$ cm$^{-2}$ and $\alpha$ from 1.2 to 2.0 results in
a $\sim 20$\% change in the conversion factor.  An upper limit to the
error in this conversion factor might be that found by using the
thermal bremsstrahlung model of PFJ93, which gives luminosity
differences up to 50\%.  For the brightest 20 X-ray sources, we
derived the 0.3--7.0~keV luminosity directly from spectral fits (see
\S3.6).

In Figure~8 we plot the cumulative luminosity function (CLF) for all
detected sources in the stacked image, and also plot separately the
CLFs of the
inner bulge (region~1), outer bulge (region~2), disk (region~3), and
bulge (regions~1+2 combined).  Histograms of the number of sources
detected against S/N peak at S/N$ = 3.5$ and fall off below this.  Clearly
we are incomplete below this level, and therefore limit our
measurements of the LFs to sources with S/N$ > 3.5$.  We fitted 
a broken power-law model to the CLFs using maximum likelihood
techniques (Cash 1979) and found the slopes listed in Table~9. 
The CLF for all sources has a break at
$(1.77\pm0.24)\times10^{37}$ \lum, with $\alpha_1=0.35\pm0.01$ before
the break and $\alpha_2=1.44\pm0.18$ after the break. This result is in
good agreement with previous \rosat\ (PFJ93) and \xmm\ (Shirey et
al. 2001) measurements of the CLF. 

The CLF for the inner bulge is significantly different, with
a break at a lower luminosity ($\sim 1.8\times10^{36}$ \lum) and a
significantly flatter distribution at the faint end. We performed a
two-sample Kolmogorov-Smirnov (K-S) test for the luminosity functions
of regions~1 and~3 and found that there is only a 3\% probability that
they are drawn from the same distribution.

In order to test if the flattening of the CLF of the inner bulge is due to
incompleteness, we performed simulations using MARX. We assumed that the
luminosity function of the sources in the inner bulge is a {\it single}\
power-law that matches the best-fit function above the break
($\alpha=0.7$) and extends to the faint end, and we generated sources at
random positions and luminosities which conformed to this
distribution.  The diffuse emission (and background) were modeled by
taking out all the detected sources, smoothing with a Gaussian and
adding Poisson noise. The sources in the simulated observation were
detected using the identical method used previously.  Detected source
counts were converted to luminosity using the same conversion
factor. We found that the luminosity function did not show flattening
below $\sim 10^{36}$ \lum. A K-S test indicates that the simulated
luminosity function is different from the actual one at a confidence level
$> 99.99$\% level. It is therefore likely that the
flattening below the break is intrinsic and not due to incompleteness.

We determined the slopes for the CLFs (above) via maximum likelihood fitting to these functions.
However, the counts in a CLF are not independent and therefore this
will underestimate the errors and may produce a biased
best estimate of the slope.  In order to more accurately estimate the
errors and slopes, we used a maximum likelihood method (e.g. Crawford,
Jauncey \& Murdoch 1970) to determine the slopes in the differential
luminosity functions (DLFs).  The slopes determined by this method are
shown in Table~9, where we have added one to the differential slopes
in order to convert them into the equivalent cumulative slopes.

The slopes above the break seem insensitive to the analysis method.
However at the faint end of the LF, below the apparent breaks, the two
methods sometimes yield significantly different slopes.  This is
particularly true in the inner bulge region, where the DLF indicates a
much smaller change in slope that we found by the CLF.  While the appearance of the LFs (Figure~9)
and the results of our MARX simulation (above) 
indicate the presence of a break to a flatter LF at the lowest
fluxes, the DLF suggests that the statistics in
the inner bulge region are insufficient to either confirm the presence or
constrain the size of a break.  However, there are sufficient counts
in the full catalog and in the disk region alone to confirm the
changes in slope using the DLF. It is worth
noting that when fitting the DLFs with a broken power-law model, the
luminosity breaks are roughly factor of 2 higher, which is consistent with
the results found by Kaaret (2002).

The black holes and neutron stars that power many of the M31 X-ray
sources have formed through the evolution of initially massive stars.
Because of this, the X-ray LF traces the history of star formation and
evolution of these massive stars in binary systems.  Breaks in the LF
may  indicate an impulsive star formation event.  As the X-ray
binaries age, their average luminosity shifts to lower values and
therefore the location of the break may be an indication of  how
long ago the star formation event occurred (Wu et al. 2002, Kilgard et
al. 2002).
Luminosity functions which do not show a break may indicate that star
formation is still occurring.  Chandra observations have measured the
breaks in the LFs of several nearby galaxies, e.g. M81 (Tennant et
al. 2001), NGC\,1553 (Blanton, Sarazin, \& Irwin 2001), NGC\,4697
(Sarazin, Irwin, \& Bregman 2001) and M83 (Soria \& Wu 2002), and these breaks have been
interpreted as evidence for impulsive star formation. 
Within M31, we find that the LF of three regions we studied has a
break at a different luminosity.  The inner bulge has this break at
the lowest luminosity, and the luminosity of the break increases
monotonically as we go out from the inner bulge.  If the breaks do
indicate the epochs of star formation events, then these events
occurred most recently in the disk of M31 and further back in time as
we move towards the nucleus of M31. 

As well as the monotonic shift in the break luminosity, there is a
monotonic shift in the slopes of the LFs.  As we move in towards the
nucleus these slopes become progressively flatter.  This is 
somewhat difficult to understand in the context of the discussion
above, because the most luminous sources would be expected to have the
shortest lifetimes.  Loss of these sources as they age would tend to
steepen the luminosity function, but we find flatter
luminosity functions in the apparently older populations.  

It is interesting to compare the LF of these three regions of M31 to
those of other galaxies.  M31 is not the first galaxy found to show a
break in its LF nor is it the first to show different LFs in different
regions; both M81 (Tennant et al. 2001) and M83 (Soria \& Wu 2002)
show similar behavior.  In cases where a single slope is fit to the LF
(i.e., where there is no clear break) the LFs of early type galaxies
and the bulges of spirals tend to be steeper ($\alpha \sim 1.7$) than
those of spiral disks and galaxies with active star forming regions
($\alpha \sim 0.8$, e.g., Prestwich 2001; Kilgard et al. 2002; Soria
\& Kong 2002).  The
opposite seems to be the case within M31: the disk has a steeper LF
than the bulge region.  We speculate that this difference may be
related to the location of the breaks in the M31 LF, which are at a
somewhat lower luminosity than those seen in other galaxies.  At these
lower luminosities we may be sampling a different class of source, and
the steepness of the LF may be due to inclusion of this new class of
faint sources rather than a loss of bright sources.  We note that
there is some evidence that we are sampling a different class of
sources as we move out from the bulge because the fraction of sources
which show variability appears lower in the disk region.  If these
sources have an intrinsically steeper LF than bright accreting
binaries, then they may be responsible for the steepening of the LFs
as we move from the bulge to the disk.  One possibility is  that these
sources are background AGN.  The presence of a class of highly cut-off
sources in the disk region (Figure 3) supports this possibility.  The
extra-galactic LF has $\alpha \sim  1.5$ at the fluxes considered in 
this paper (see Rosati et al. 2002), so a larger contribution from
background sources could explain some of the steepening we observe.
A search for optical counterparts to these cut-off sources could test
this possibility. 

\section{Summary}

By using a stacked image (39.7 ks) of M31 from \chandra\ ACIS-I data
taken between 1999 and 2001, we have detected 204 X-ray sources in the
central $\sim 17'\times17'$ region of M31, with luminosity above $\sim
1.6\times 10^{35}$ \lum. Of these 204 sources, we identified 22
globular clusters, 2 supernova remnants, 9 planetary nebula, 9 SSSs and 1
background object.  We suggest that an additional 4 source are normal
stars based on both their positional coincidence and soft spectra,
and another $\sim 30$ sources may be background AGN based on deep
field observations and their hard (possibly absorbed) spectra.
We do not detect any OB~associations or stellar nova. 

We find 10~positional matches with M31 planetary nebula, but expect
that $\sim 2$ of these may be random coincidences.  The X-ray
luminosity of these sources is an astounding 5 to 7 orders of
magnitude higher than planetary nebula in our Galaxy.  We suggest that
these may not be planetary nebula at all, but analogs to GX 1+4 (i.e.,
symbiotic stars with a neutron star primary).  Optical spectroscopy
of these sources could confirm or refute this suggestion.

About 50\% of all the detected sources are variable on time scales of
months. We also found 13 transients, corresponding to $\sim 7$\% of
the total population.  These transients show a variety of lightcurves,
including the classical ``FRED'' (fast rise and exponential decay) with an
e-folding decay time of $\sim 30$ days, an outburst lasting $>1$~year, and a
possible periodicity of $\sim 200$~days in one case.  The first of these
behaviors is analogous to X-ray nova in our Galaxy, the last
reminiscent of the Galactic globular cluster source 4U~1820-30.

The median energy spectrum of point sources can be represented by a
single power-law with a photon index of $\sim 1.7$.  There are 9
sources with hardness ratios indicative of SSSs.  The spectra of 12
sources are shown to be variable.  The spectral variations and
luminosity of the brightest of these sources is reminiscent of Galactic Z
sources on the normal branch.

The luminosity function of all the X-ray point sources is consistent with
the findings of \rosat\ and \xmm\ observations (PFJ93, Su97 and Shirey
et al. 2001), with a break at $\sim 1.7\times10^{37}$ \lum\ above
which the function steepens. We also found that the luminosity
functions of three separate regions (roughly
corresponding to the inner bulge, outer bulge and disk) are
different. In particular, the inner bulge shows a break near $10^{36}$
\lum and it shifts monotonically to higher luminosities when moving
outward from the nucleus to the outer bulge and disk. In addition, the
slopes become steeper, indicating that the star formation and
evolution histories might be different for the bulge and disk sources.
Hence, our \chandra\
observations reveal different star formation histories even within the
central $\sim 17'\times17'$ ($\sim 3.9$ kpc) region of M31. Future
\chandra\ and \xmm\ observations along the disk will definitely
improve our knowledge of the star formation history of the whole galaxy.  

\begin{acknowledgements}
We are grateful to Kinwah Wu, Andrea Prestwich and Phil Kaaret for stimulating
discussion and comments.  We thank Phil Kaaret for providing the code
for the broken
power-law model. We also thank the referee for a very clear and
helpful report on our manuscript. 
AKHK was supported by a Croucher
Fellowship. MRG acknowledges the support of
NASA LTSA Grant NAG5-10889 and NASA Contract NAS8-39073 to
the CXC. The HRC GTO program is supported by NASA Contract
NAS-38248. This research has made use of the SIMBAD database, operated
at CDS, Strasbourg, France, and the
NASA/IPAC Extragalactic Database (NED) which is operated by the Jet
Propulsion Laboratory, Caltech, under contract with NASA.

\end{acknowledgements}

\clearpage

\begin{deluxetable}{ccc}
\tablecaption{Journal of \chandra\ Observations}
\tablewidth{0pt}
\tablehead{Date & Obs ID & Exposure time (s)}
\startdata
Oct 13 1999 & 303 & 8830\\
Dec 11 1999 & 305 & 4129\\
Dec 27 1999 & 306 & 4132\\
Jan 29 2000 & 307 & 4113\\
Feb 16 2000 & 308 & 4012\\
Jul 29 2000 & 311 & 4894\\
Aug 27 2000 & 312 & 4666\\
Jun 10 2001 & 1583 & 4903\\
\enddata
\end{deluxetable}

\clearpage

\input{table2.tex}

\begin{deluxetable}{llcccc}
\tablewidth{0pt}
\tablecaption{Summary of Source IDs}
\tablehead{Object  &Catalogs 	& Searching & $N_m$ & $N_{acc}$ & $N_{true}$\\
type    &         		& radius ($''$) &   &  &}
\startdata
X-ray 	& \rosat\ HRI (PFJ93) 	& 6 & 77 & 9.5 & 67\\
GC 	& Ba87, Magnier (1993), and Barmby (2001) & 3 & 23& 2.25 & 22\\
SNR 	& DO80, BW83, and Ma95 	& 10& 2 & 1  & 2\\
PN 	& Ford78, Ci89		& 3 &11 & 5  & 9\\
OB Assoc. & Magnier et al. 1993	& 3 & 0 & 0.25 & 0\\
Nova   	& IAUC 			& 3 & 0 & 0.25 & 0\\
Extragalactic & NED and SIMBAD 	& 3 & 1 & 1    & 1\\
Stars	& Ha94 and SIMBAD 	& 0.8 & 29 & 21.25 & 4\\
\enddata
\tablecomments{$N_m$: Number of all possible matches; $N_{acc}$:
Number of matches by accident; $N_{true}$: likely number of true
matches.}
\tablerefs{Ba87: Battistini et al. 1987; DO80: d'Odorico, Dopita,
\& Benvenuti (1980); BW83: Braun \& Walterbos (1993); Ma95: Magnier et
al. (1995); Ford78: Ford \& Jacoby (1978); Ci89: Ciardullo et
al. (1989); Ha94: Haiman et al. (1994)}

\end{deluxetable}

\input{table4.tex}

\input{table5.tex}

\begin{deluxetable}{ccc}
\tablewidth{0pt}
\tablecaption{Average Hardness Ratios}
\tablehead{Region & HR1 & HR2}
\startdata
Inner bulge & $0.30\pm0.03$ & $0.00\pm0.03$\\
Outer bulge & $0.32\pm0.05$ & $0.11\pm0.04$\\
Disk & $0.48\pm0.03$ & $0.30\pm0.04$\\
\enddata
\end{deluxetable}

\begin{deluxetable}{cccccccc}
\tablewidth{0pt}
\tablecaption{Variable sources in M31}
\tablehead{Region & Detected  & Number of & Fraction & Number of &
Fraction & Number of & Fraction \\
    &  sources & variables & & spectral variables& & transients & }
\startdata
Inner bulge & 33 & 24 & 73\% & 3 & 9\% &5 & 15\%\\
Outer bulge & 59 & 34 & 58\% & 7& 12\% & 5 & 8.5\%\\
Disk & 106 & 42 & 39\% & 2 & 2\% & 3& 2.8\%\\
\enddata

\tablecomments{6 sources are excluded in the disk region as they were in the field
in only one observation; the actual total detected sources in this region is 112.}
\end{deluxetable}

\begin{deluxetable}{lccccccccc}
\tablewidth{0pt}
\tabletypesize{\scriptsize}
\tablecaption{Spectral fits to the 20 brightest sources}
\tablehead{
& \multicolumn{4}{c}{Power-law} & &\multicolumn{4}{c}{Raymond-Smith\,$^a$} \\
\cline{2-5} \cline{7-10}\\
\colhead{Source} & \colhead{$N_H$} & \colhead{$\alpha$} &
\colhead{$\chi^2_{\nu}/dof$} & \colhead{Flux\,$^b$}& & \colhead{$N_H$}
&\colhead{$kT_{RS}$} & \colhead{$\chi^2_{\nu}/dof$} &\colhead{Flux\,$^b$}\\
\colhead{ID} & \colhead{($10^{21}$ cm$^{-2}$)} & & & \colhead{} & &\colhead{($10^{21}$
cm$^{-2}$)} & \colhead{(keV)}& & \colhead{} 
}
\startdata
r1-1 & $2.3^{+0.7}_{-0.6}$ & $1.7\pm0.2$ & 0.8/36 & 1.38 &&
       $1.9\pm0.5$&$7.4^{+3.3}_{-1.9}$& 0.8/36& 1.32\\
r1-5 & $2.3^{+0.7}_{-0.6}$ & $1.6\pm 0.2$ & 0.9/31 & 1.25 && $1.9\pm0.6$&$10.9^{+14.7}_{-4.2}$&0.9/31&1.22\\
r2-3 & $2.7^{+0.6}_{-0.9}$ & $2.5^{+0.4}_{-0.3}$ & 1.2/31 & 0.49
       &&$1.0\pm0.5$&$3.3^{+1.1}_{-0.8}$&1.5/13 & 0.37\\
r2-5 & $0.5^{+0.8}_{-\infty}$ & $1.1^{+0.2}_{-0.3}$ & 0.9/11 & 0.58 &&
       $0.8^{+0.7}_{-0.5}$&$64.0^{+\infty}_{-46.5}$& 1.0/11& 0.52\\
r2-6 & $1.8^{+1.0}_{-0.8}$ & $1.9^{+0.4}_{-0.3}$ & 0.7/12 & 0.45 &&
       $0.8\pm0.6$&$7.3^{+10.5}_{-2.7}$&0.9/12 & 0.44\\
r2-11 & $1.8^{+0.7}_{-0.6}$ & $1.8\pm0.2$ & 1.4/30 & 0.96 &&
       $1.0\pm0.5$&$7.5^{+3.8}_{-1.9}$& 1.5/30& 0.94\\
r2-13 & $1.8^{+0.8}_{-0.6}$ & $1.5\pm0.2$ & 1.0/33 & 1.31 &&
       $1.6\pm0.6$&$14.8^{+28.3}_{-6.2}$& 1.0/33 & 1.29\\
r2-26 & $1.9^{+0.7}_{-0.5}$ & $1.0\pm0.1$ & 1.0/62 & 3.42 &&
       $3.2\pm0.5$&$64.0^{+\infty}_{-20.3}$& 1.3/62 & 3.07\\
r2-26$^d$ & $3.2^{+1.7}_{-1.4}$ & $1.6^{+0.3}_{-0.2}$ & 0.8/22 & 3.07 &&
       $2.6^{+1.4}_{-1.1}$ & $9.0^{+13.0}_{-3.5}$ & 0.8/22& 3.00 \\
r2-32 & $< 0.5$ & $1.3^{+0.2}_{-0.1}$ & 0.6/12 & 0.49 && 0.0\,$^c$&
       $48.6^{+\infty}_{-51.9}$ & 0.7/12 & 0.48\\
r2-34 & $1.5^{+1.0}_{-0.8}$ & $1.9\pm0.3$ & 1.5/13 & 0.60 &&
       $0.7\pm0.6$&$6.0^{+4.7}_{-1.8}$& 1.5/13 &0.57\\
r2-35 & $0.9^{+0.7}_{-0.6}$ & $1.4\pm0.2$ & 1.0/22 & 0.89 &&
       $0.7\pm0.5$& $24.4^{+\infty}_{-16.5}$&1.0/22 & 0.88\\
r3-15 & $0.9\pm0.5$ & $1.4\pm0.2$ & 0.9/40 & 1.62 &&
       $0.8^{+0.4}_{-0.2}$& $30.4^{+\infty}_{-18.3}$ & 0.9/40 & 1.60\\
r3-16 & $2.5^{+0.2}_{-0.1}$ & $1.9^{+0.5}_{-0.3}$ & 1.0/12 & 0.50 &&
       $1.4^{+0.8}_{-0.6}$& $6.2^{+5.8}_{-1.9}$& 1.1/12& 0.46\\
r3-22 & $2.6\pm0.1$ & $2.0\pm0.3$ & 0.7/13 & 0.51 &&
       $1.4^{+0.8}_{-0.6}$& $4.6^{+2.5}_{-1.3}$& 0.7/13 & 0.45\\
r3-39 & $2.6\pm0.1$ & $1.8\pm0.3$ & 0.7/18 & 0.80 && $1.8\pm0.7$&
       $7.1^{+6.4}_{-2.3}$ & 0.8/18 & 0.76\\
r3-42 & $5.9\pm0.2$ & $2.5^{+0.5}_{-0.4}$ & 0.5/11 & 0.65 &&
       $3.5^{+1.2}_{-0.9}$& $3.4^{+1.5}_{-0.9}$& 0.6/11& 0.46\\
r3-44 & $1.6\pm0.8$ & $1.5\pm0.2$ & 1.2/23 & 0.90 &&
       $1.2\pm0.6$&$13.2^{+43.2}_{-5.9}$& 1.1/23& 0.87\\
r3-45 & $3.5\pm1.0$ & $2.8\pm0.4$ & 1.0/19 & 0.79 &&
       $0.9\pm0.5$&$3.1^{+0.9}_{-0.6}$& 1.6/19& 0.51\\
r3-52 & $8.5^{+0.2}_{-0.1}$ & $3.0\pm0.3$ & 0.8/26 & 2.05&&
       $5.4^{+1.3}_{-0.9}$&$2.4^{+0.5}_{-0.4}$& 1.1/26& 1.05\\
r3-61 & $4.7\pm0.2$ & $1.8\pm0.3$ & 1.0/13 & 0.70 &&
       $3.8\pm1.0$&$6.9^{+7.8}_{-2.3}$& 0.9/13 &0.67\\
\enddata
\normalsize
\tablecomments{All quoted uncertainties are 90\% confidence.}
\tablenotetext{a}{Fixed at solar abundence}
\tablenotetext{b}{Unabsorbed flux in 0.5--10 keV ($10^{-12}$ \flux)}
\tablenotetext{c}{$N_H$ hit the minimum value of 0 allowed by XSPEC}
\tablenotetext{d}{Pile-up corrected}
\end{deluxetable}

\begin{deluxetable}{ccccccc}
\tablewidth{0pt}
\tabletypesize{\small}
\tablecaption{Luminosity functions of M31}
\tablehead{
Region &\multicolumn{3}{c}{Cumulative\,\tablenotemark{a}}
&&\multicolumn{2}{c}{Differential\,\tablenotemark{b}} \\
\cline{2-4} \cline{6-7}\\

\colhead{}&\colhead{$\alpha_1$\,\tablenotemark{c}} &
       \colhead{$\alpha_2$\,\tablenotemark{d}} & \colhead{Break}
       &&\colhead{$\alpha_1$\,\tablenotemark{c}}&\colhead{$\alpha_2$\,\tablenotemark{d}}\\
&                &                & ($\times10^{37}$ \lum) &}
\startdata
Inner bulge & $0.12\pm0.03$ & $0.67\pm0.08$ &$0.18\pm0.08$& &$0.88^{+0.57}_{-0.40}$
       & $0.73^{+0.27}_{-0.23}$\\
Outer bulge & $0.22\pm0.03$ & $0.89\pm0.13$ &
$0.69^{+0.20}_{-0.15}$  && $0.58^{+0.20}_{-0.15}$ & $0.78^{+0.35}_{-0.26}$\\
Bulge & $0.23\pm0.02$ & $0.86\pm0.09$ &
$0.56\pm0.10$ && $0.58^{+0.15}_{-0.13}$ & $0.80^{+0.24}_{-0.21}$\\
Disk & $0.41\pm0.02$ & $1.86\pm0.40$ &
$2.10\pm0.39$ && $0.55^{+0.10}_{-0.09}$ &
       $1.93^{+0.94}_{-0.72}$\\
Integrated & $0.35\pm0.01$ & $1.44\pm0.18$ &
$1.77\pm0.24$ && $0.50^{+0.08}_{-0.06}$& $1.58^{+0.47}_{-0.40}$\\
\enddata
\tablecomments{All quoted uncertainties are 90\% confidence.}
\tablenotetext{a}{$N(>L)=K_1L^{-\alpha}$ where $K_1$ is normalization}
\tablenotetext{b}{$\frac{dN}{dL}=K_2L^{-(\alpha+1)}$ where $K_2$ is normalization} 
\tablenotetext{c}{Power-law slope below the break}
\tablenotetext{d}{Power-law slope above the break}
\end{deluxetable}

\begin{figure}[t]
\begin{center}
\psfig{file=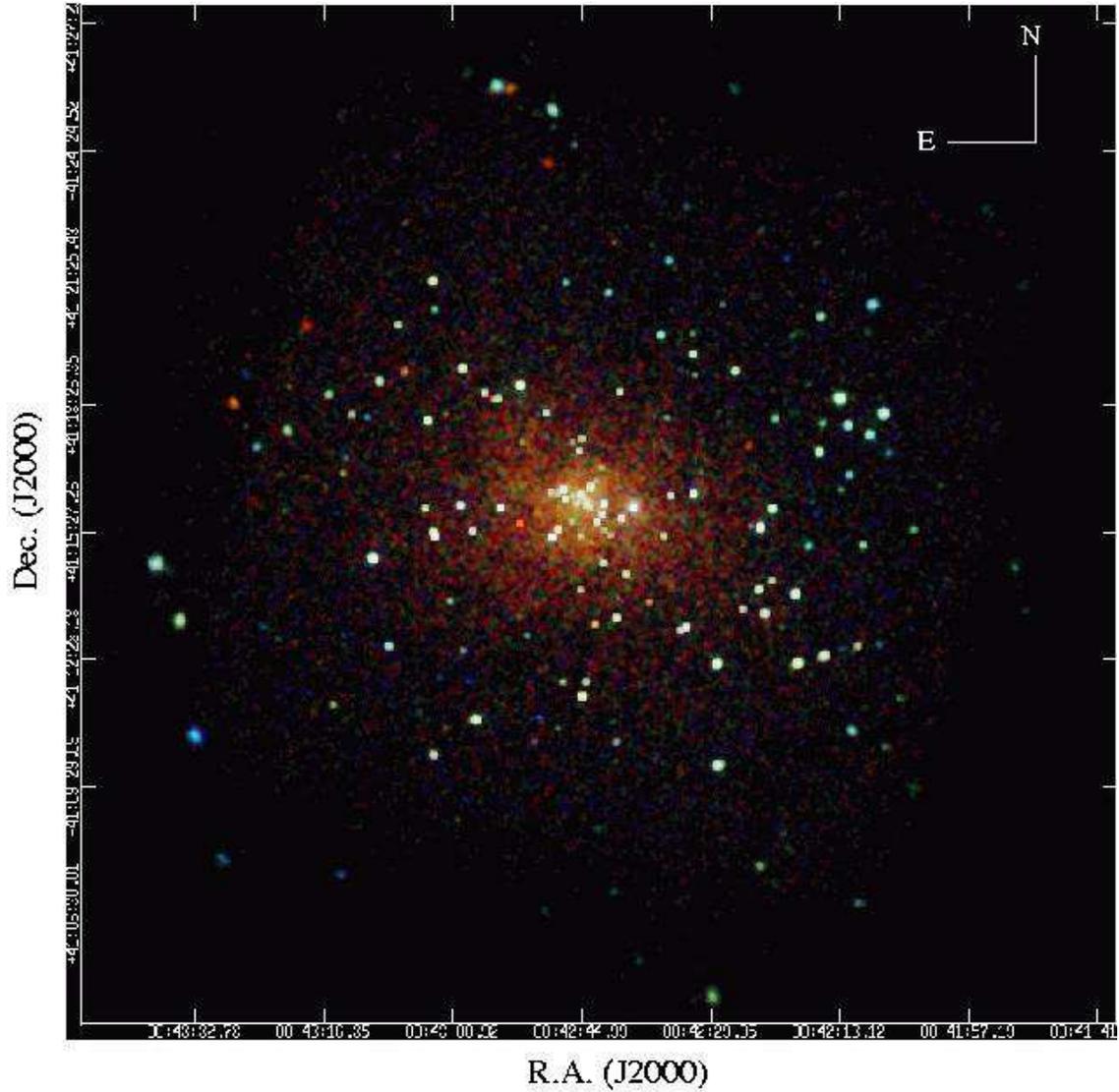,height=15cm,width=15cm}
\end{center}
\caption{Stacked ``true color'' \chandra\ ACIS-I image (39.7 ks) of the central
$\sim 17'\times17'$ region of M31. This image was constructed from the
soft (red; 0.3--1 keV), medium (green; 1--2 keV) and hard (blue; 2--7
keV) energy bands. The pixel size is $1.96''$ and the image has been
smoothed with a $1.96'' \sigma$ Gaussian function. The brightness scale
is arbitrary in order to show the details.}
\end{figure}

\begin{figure}[t]
\begin{center}
\psfig{file=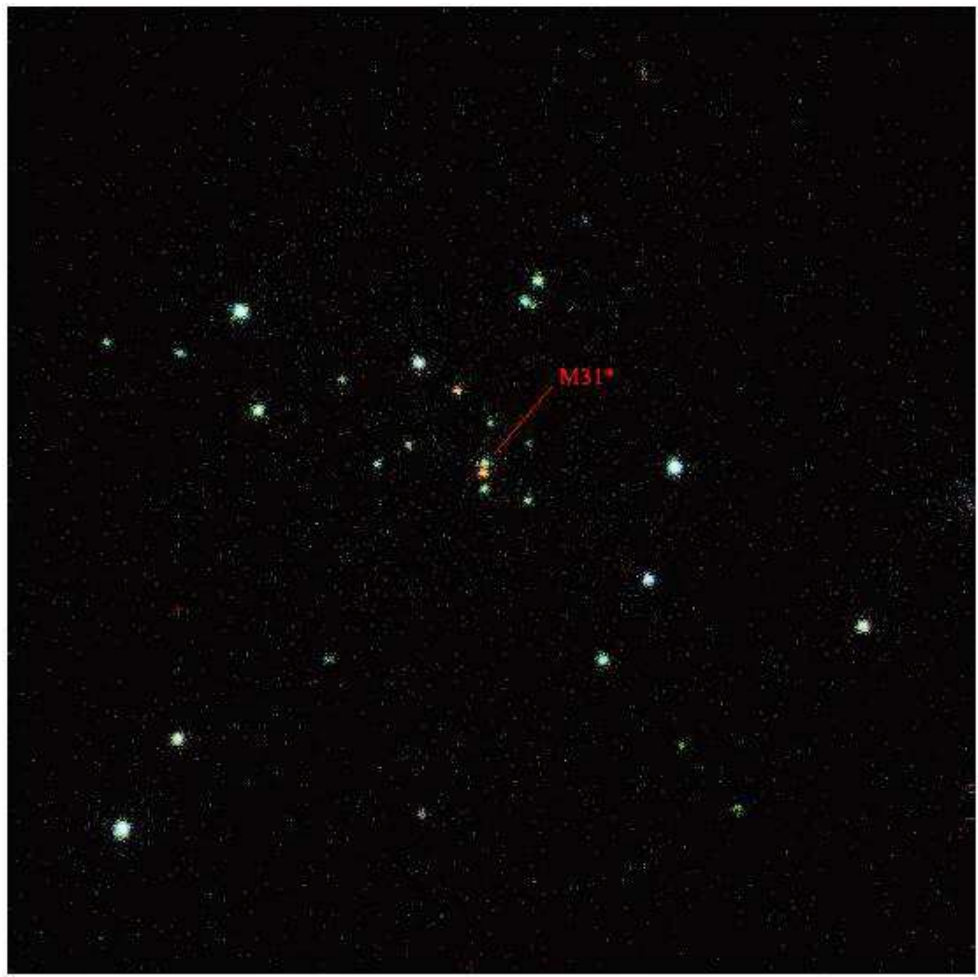,height=13cm,width=13cm}
\end{center}
\caption{Stacked ``true color'' \chandra\ ACIS-I image of the central
$\sim 2'\times2'$ region (``region~1'') of M31. The color
representation is the same as Figure 1 and the pixel size is
$0.123''$. M31$^*$ candidate is marked.}
\end{figure}

\begin{figure}[t]
\begin{center}
\psfig{file=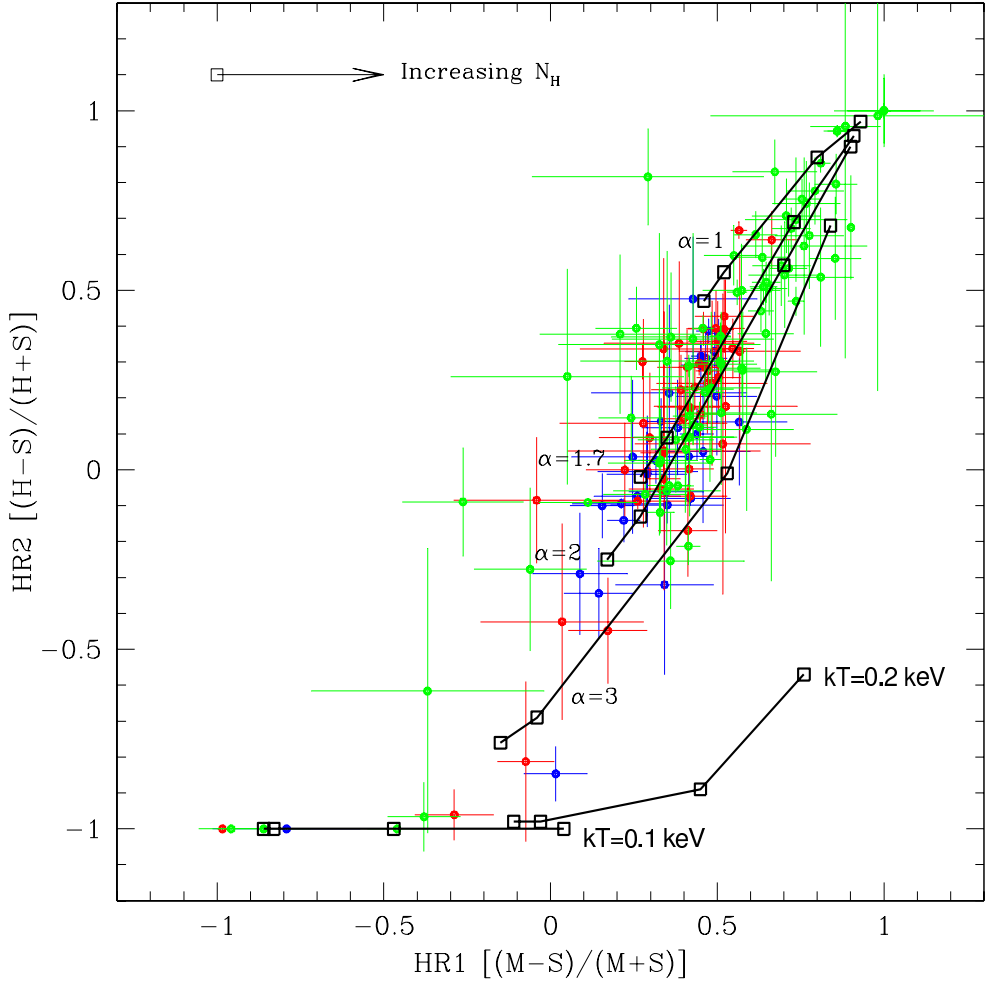,height=13cm,width=13cm}
\end{center}
\caption{Color-color diagram for all sources with more than 20
counts. Color symbols represent different regions of the field:
region~1 (blue); region~2 (red) and region~3 (green). Also plotted
are the estimated
hardness ratios estimated from different spectral models. From top to bottom:
power-law model with $\alpha$ of 1, 1.7, 2 and 3, and blackbody model
with $kT$ of 0.2 keV and 0.1 keV. For each model, $N_H$ varies from
the left from $5\times10^{20}$, $10^{21}$, $5\times10^{21}$ and
$10^{22}$. r-1, r-2 and r-3 represents the inner bulge, outer bulge
and disk, respectively (see \S\,3.2 for details).}
\end{figure}

\begin{figure}[t]
\begin{center}
\psfig{file=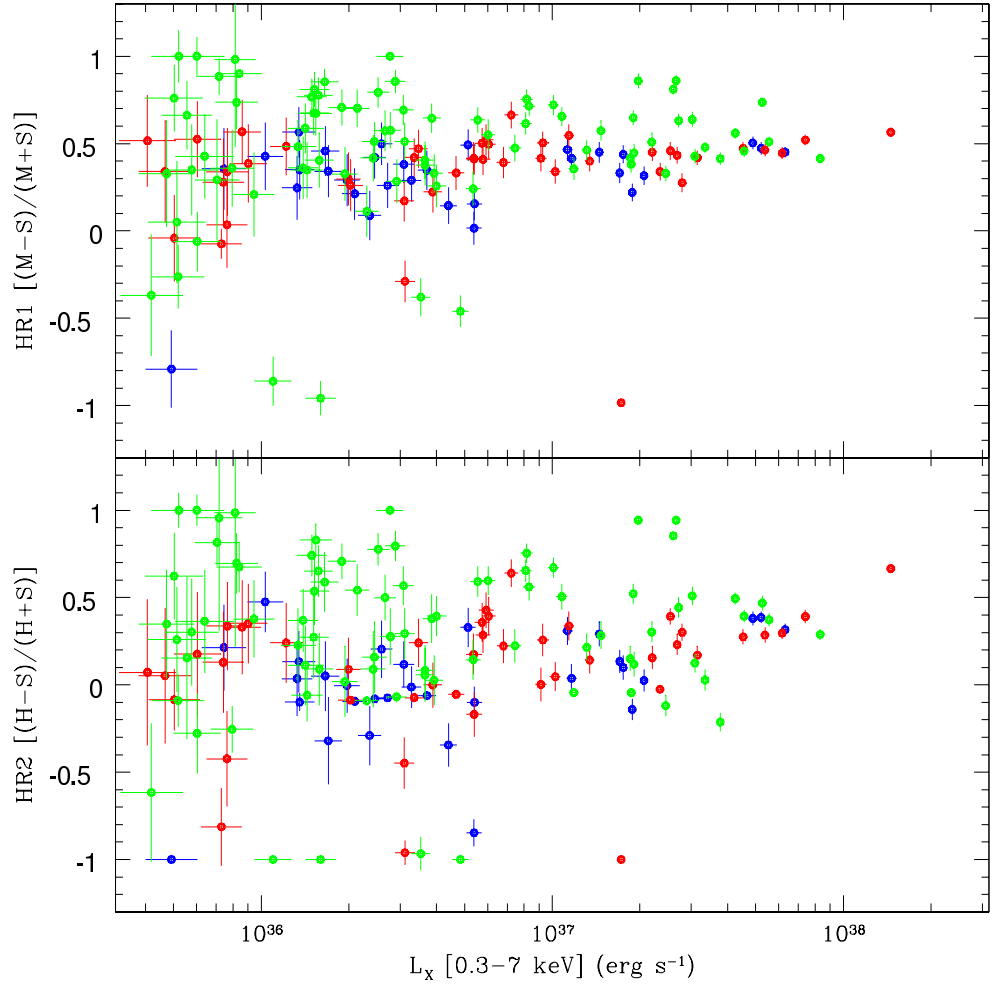,height=13cm,width=13cm}
\end{center}
\caption{Hardness-intensity diagram for all sources with more than 20
counts. Color symbols are the same as Fig. 3.}
\end{figure}

\begin{figure}[t]
\begin{center}
\psfig{file=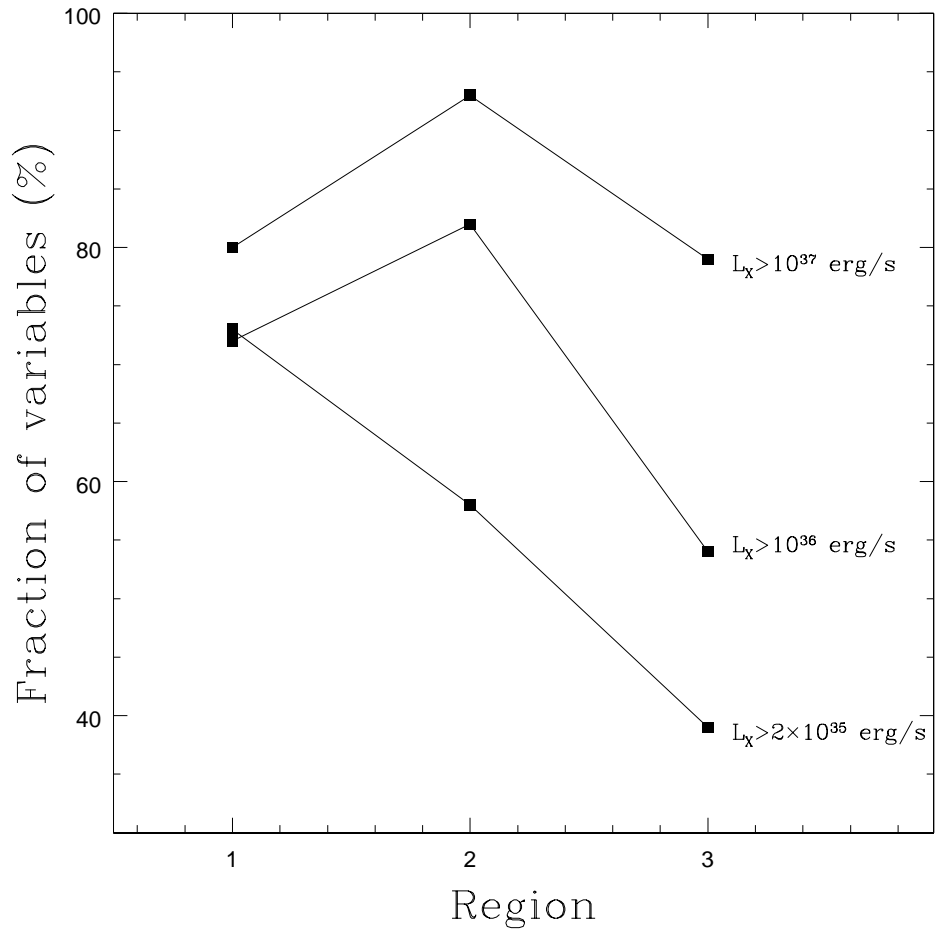,height=13cm,width=13cm}
\end{center}
\caption{Fraction of variables in different regions.}
\end{figure}

\begin{figure}[t]
\begin{center}
\psfig{file=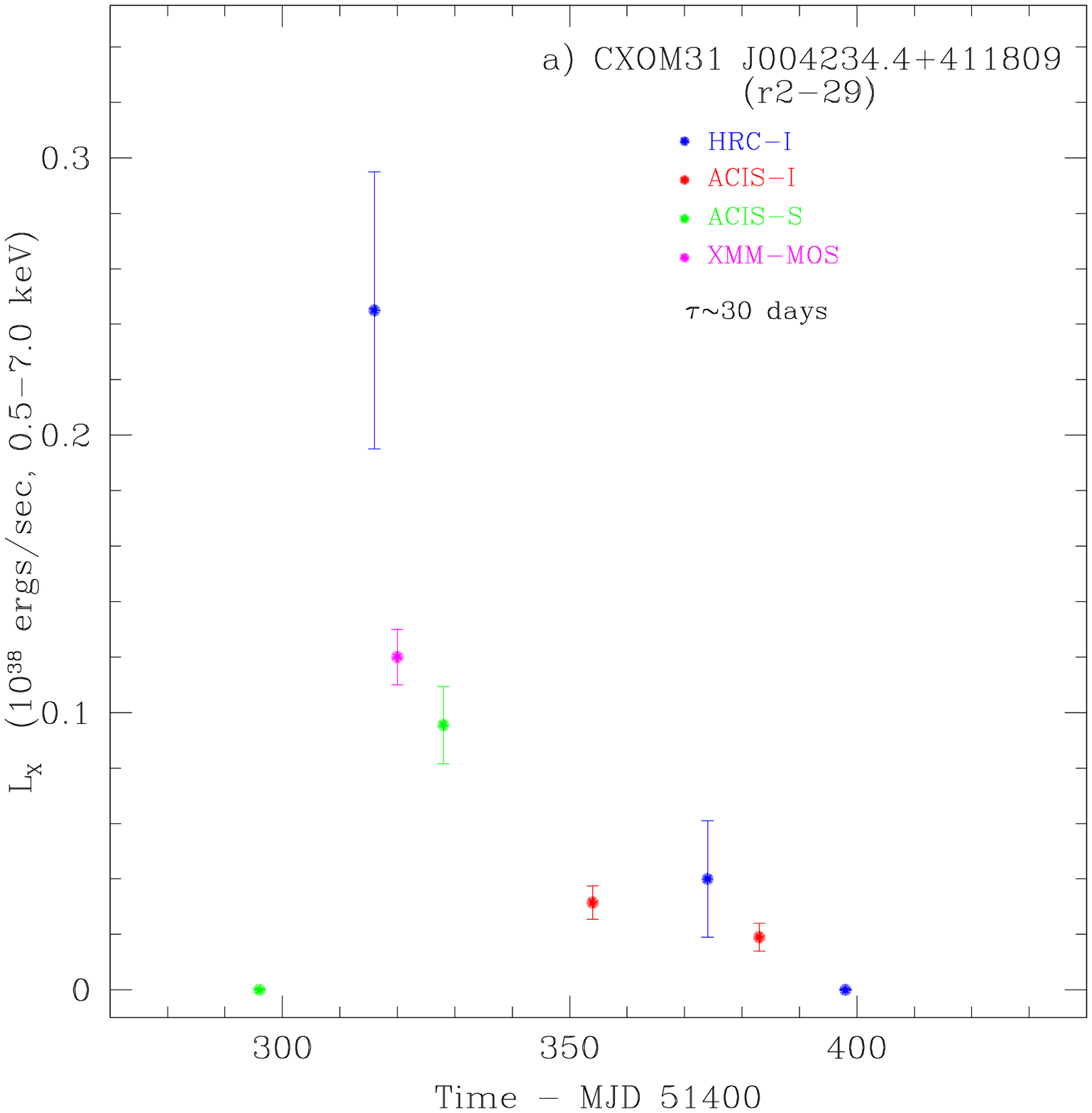,height=8cm,width=8cm}
\psfig{file=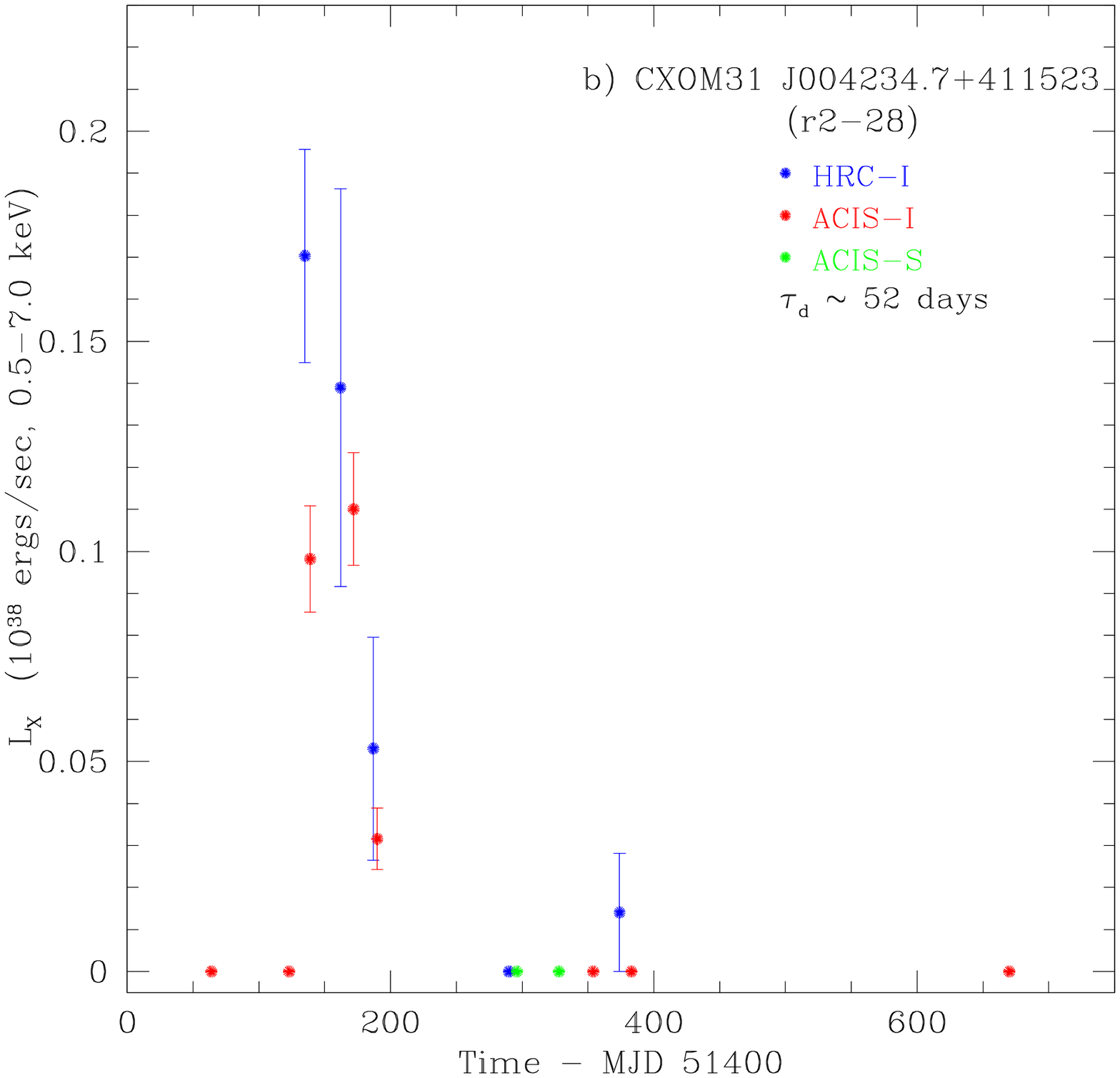,height=8cm,width=8cm}
\psfig{file=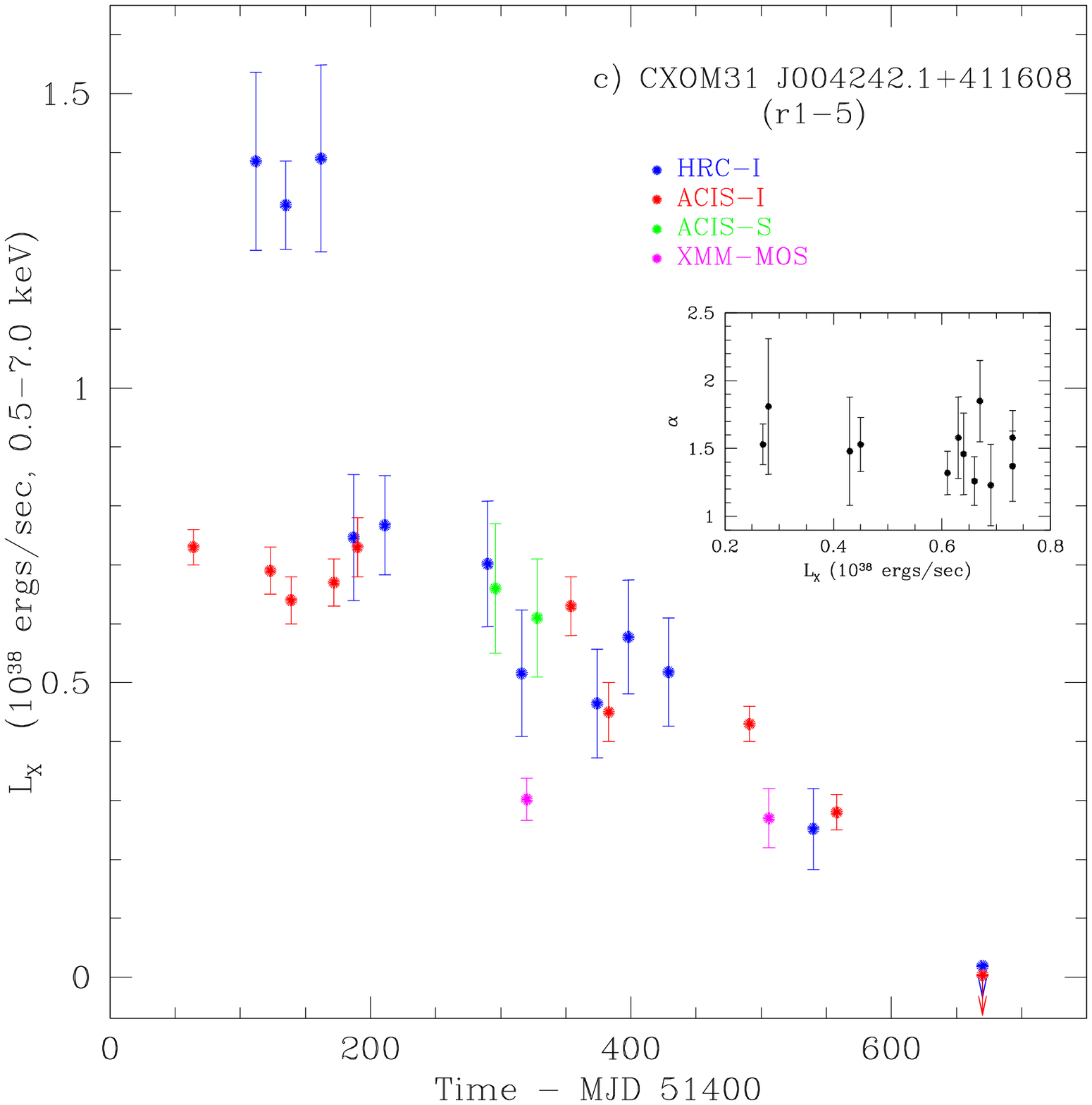,height=8cm,width=8cm}
\psfig{file=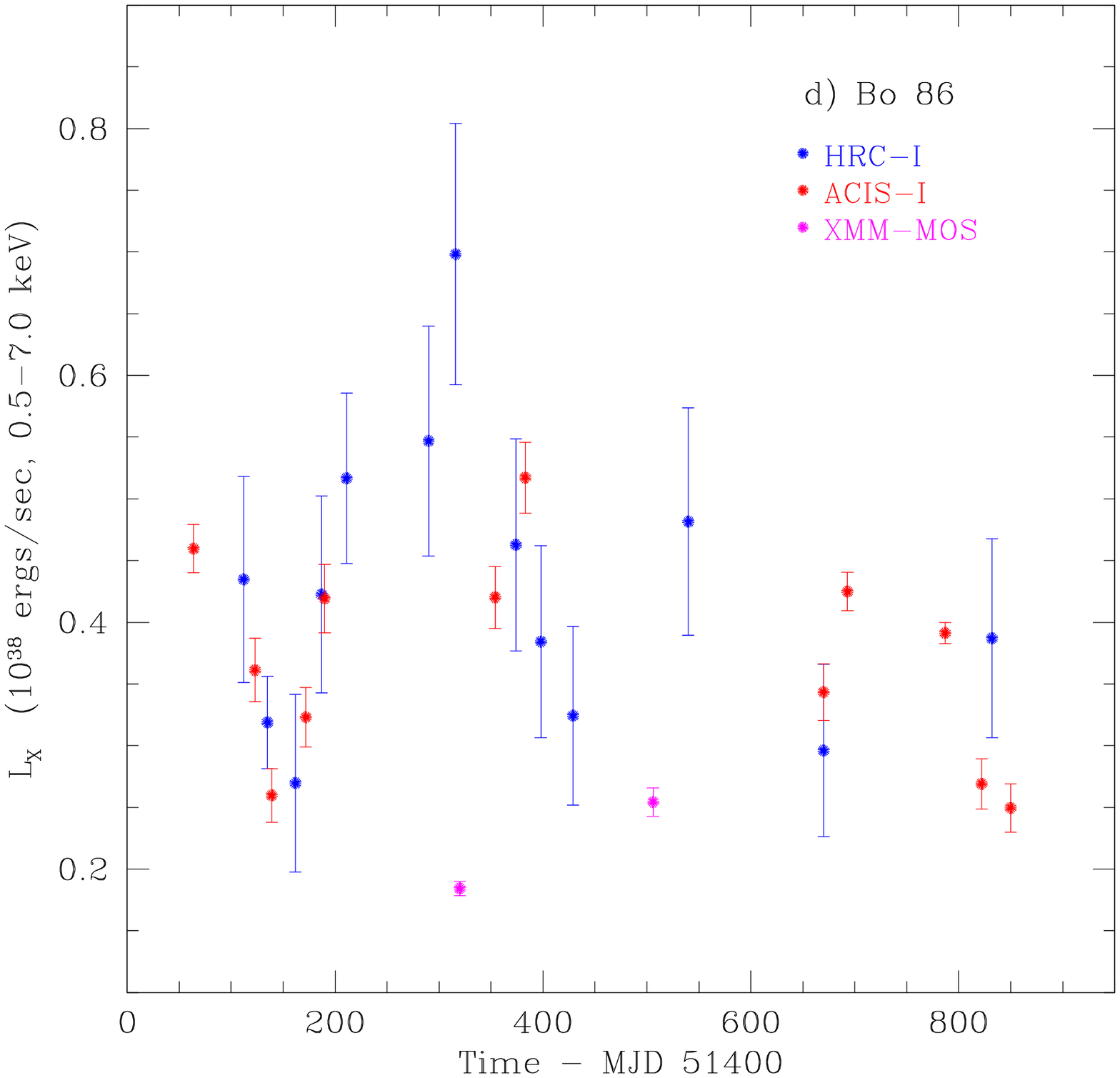,height=8cm,width=8cm}
\end{center}
\caption{Light curves of 3 bright transients and globular cluster
Bo\,86 as seen in the past two years with \chandra. Observations from
HRC-I, ACIS-S and XMM-MOS are included. The inset of r1-5 shows that
the energy spectrum during the whole outburst is consistent with a
power-law with slope of $\sim 1.5$ (see also Trudolyubov, Borozdin, \&
Priedhorsky 2001). It is worth noting that the three HRC-I data points
during the early stage of the outburst
indicate that the source underwent flarings or state transitions.}
\end{figure}

\begin{figure}[t]
\begin{center}
\psfig{file=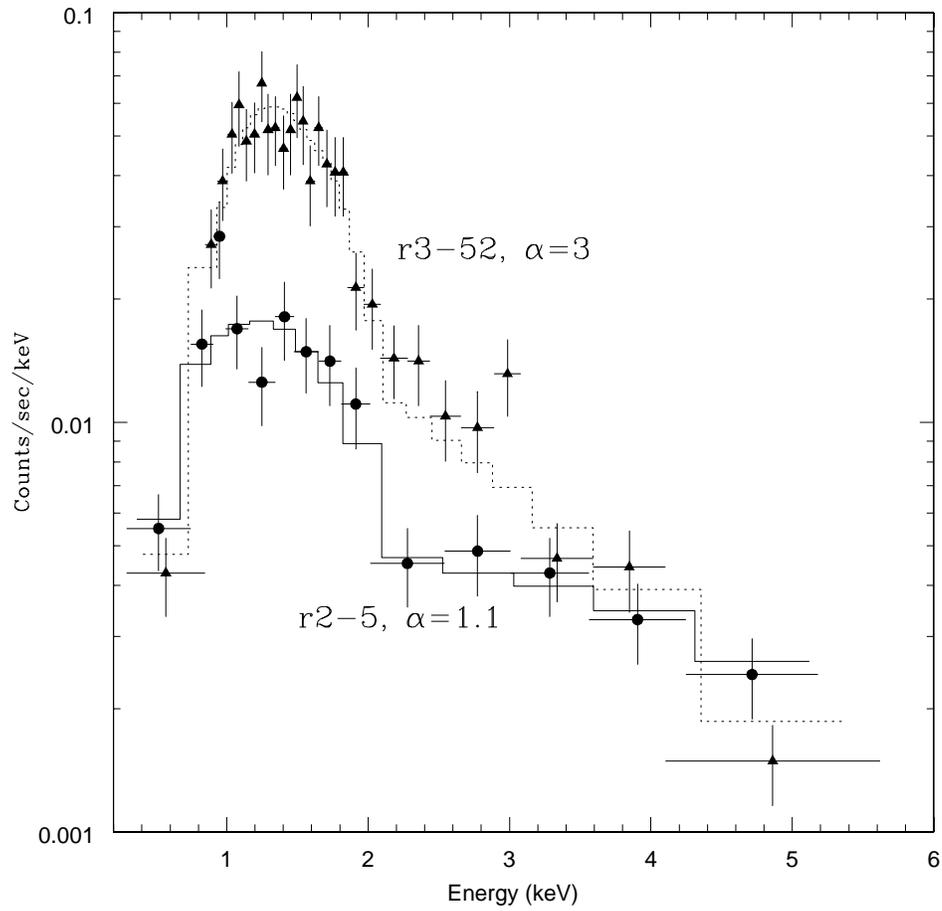,height=13cm,width=13cm}
\end{center}
\caption{The softest (r3-52; triangles) and the hardest (r2-5;
circles) energy spectra from the 20
brightest X-ray sources in the first observation.}
\end{figure}

\begin{figure}[t]
\begin{center}
\psfig{file=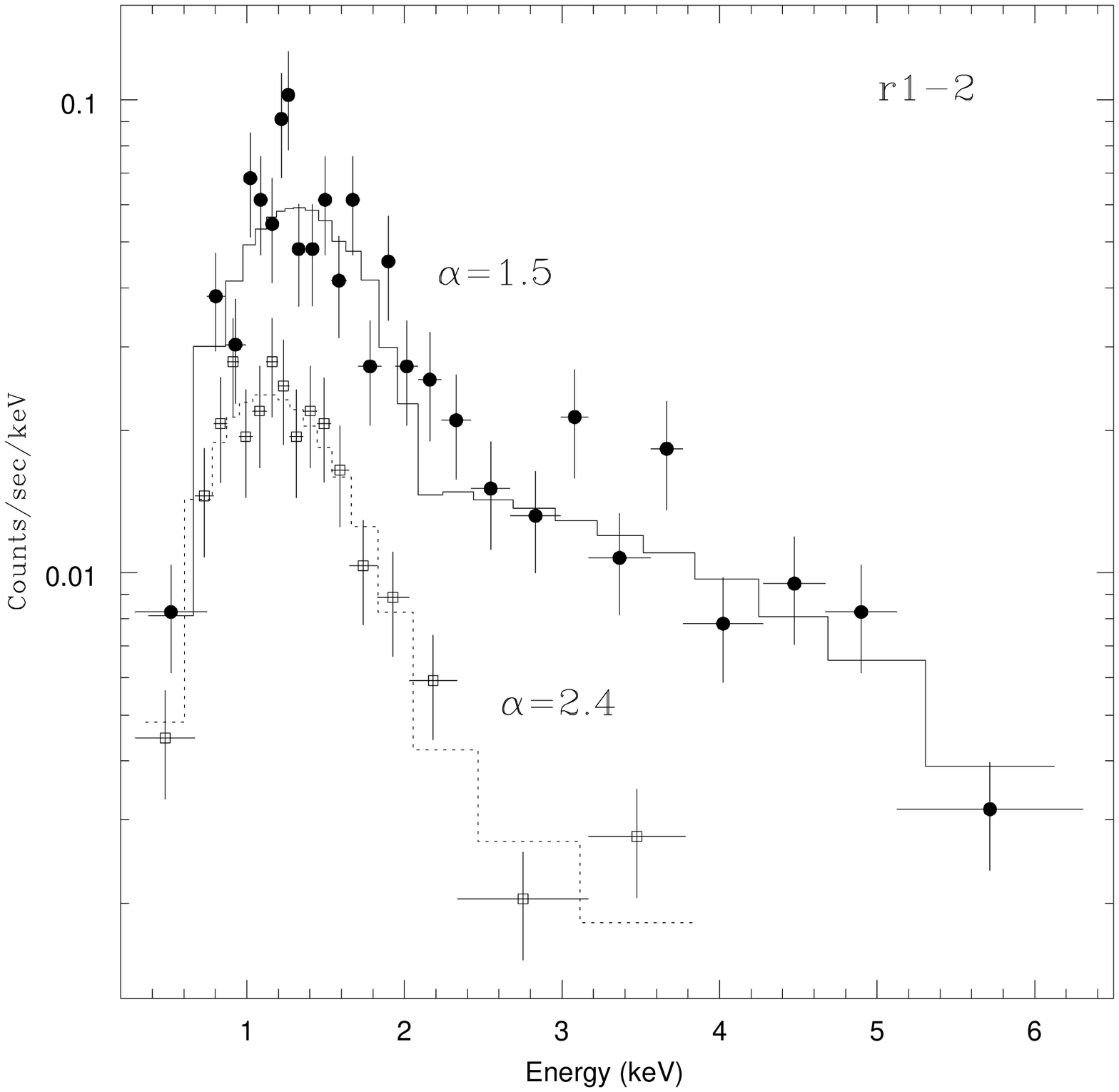,height=9cm,width=8.1cm}
\psfig{file=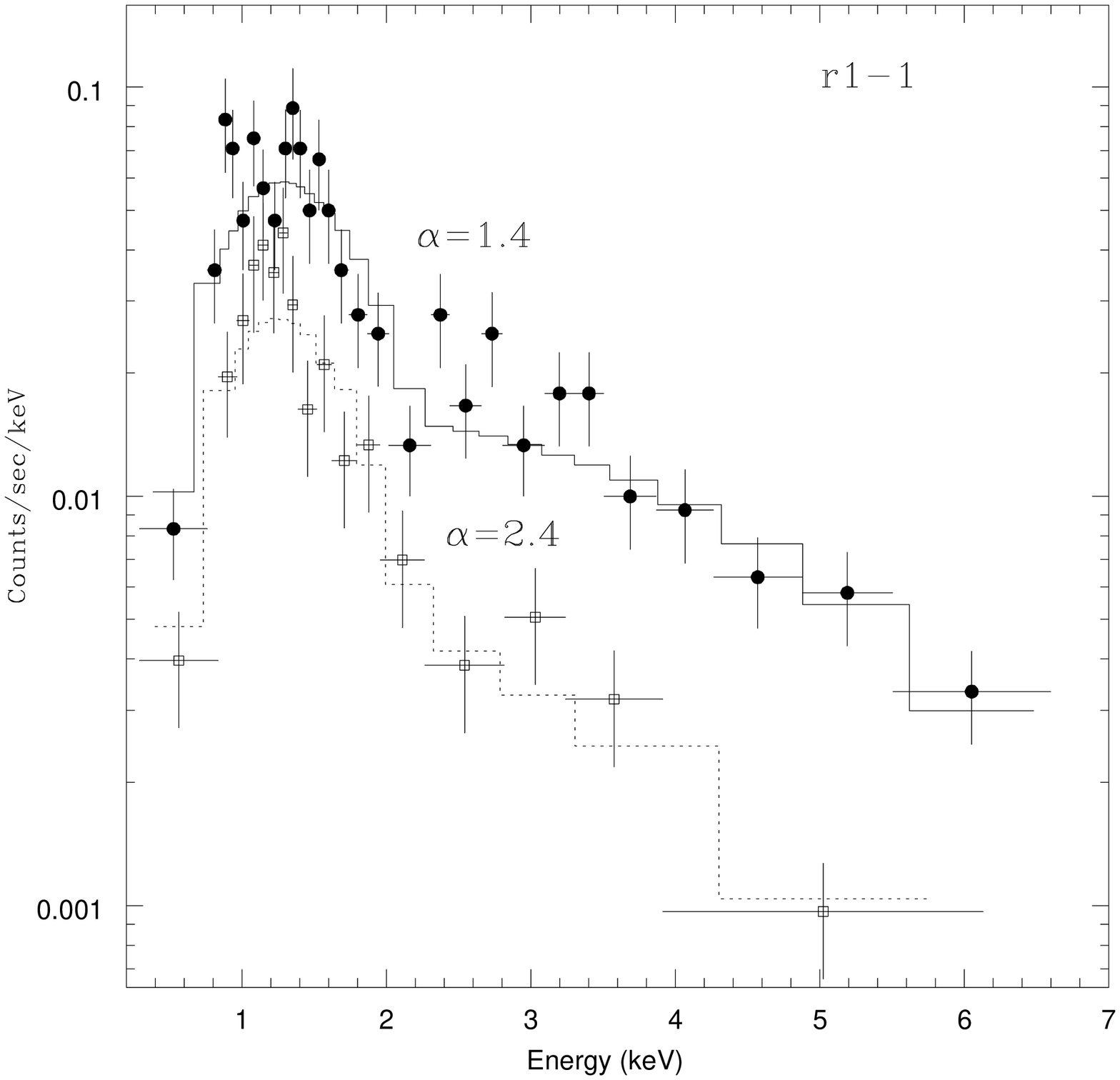,height=9cm,width=8.1cm}
\end{center}
\caption{Spectral changes of two of the sources. They change from
lower luminosity at softer state to higher luminosity with harder state.}
\end{figure}

\begin{figure}[t]
\begin{center}
{\rotatebox{-90}{\psfig{file=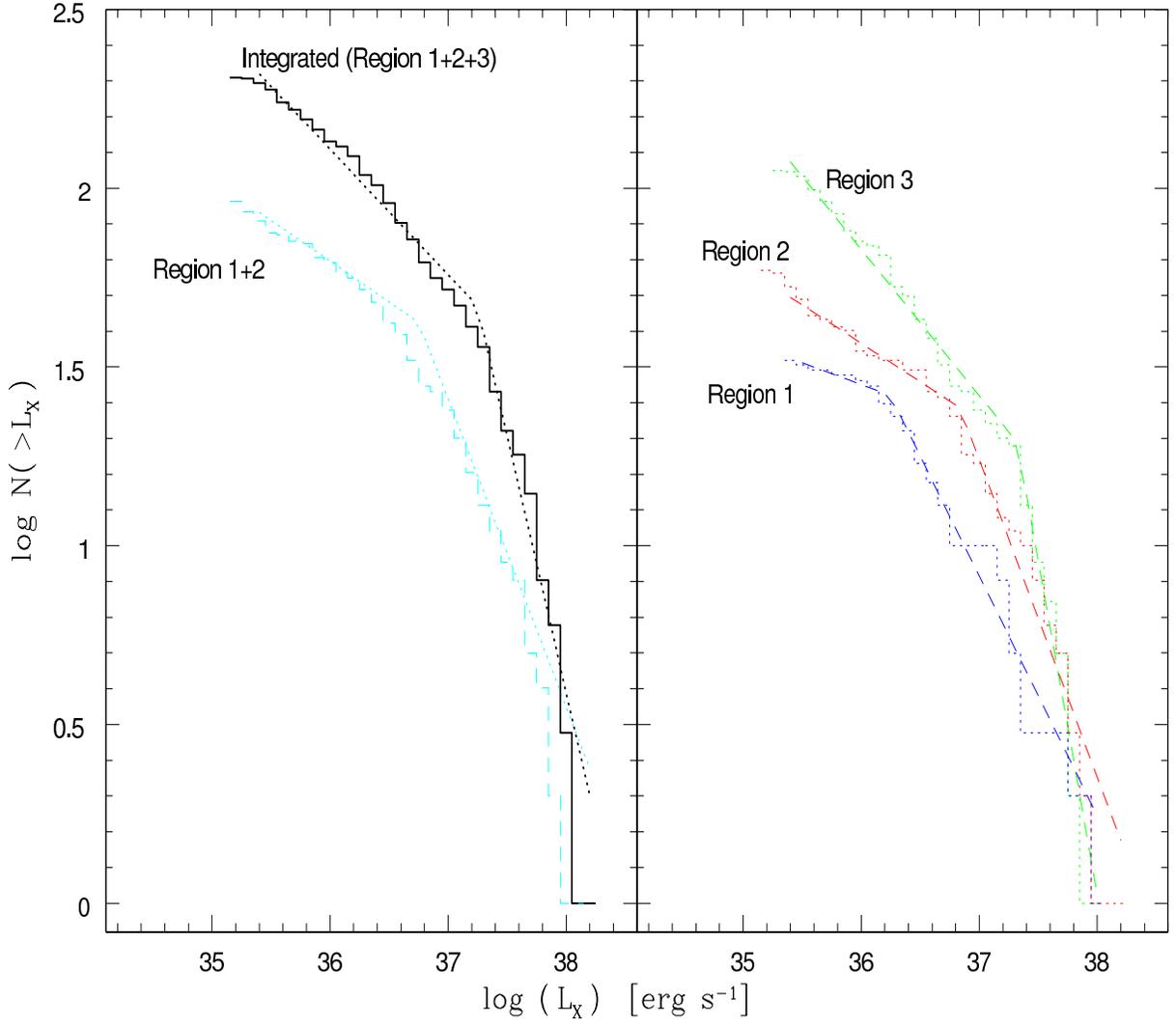,height=16cm,width=14cm}}}
\end{center}
\caption{Left: Luminosity functions for all sources (Regions 1+2+3)
and bulge (Regions 1+2).
Right: Luminosity functions for inner bulge (Region 1), outer bulge (Region 2) and
disk (Region 3).}
\end{figure}

\end{document}

%% file: table2.tex
\begin{deluxetable}{lcccccccccl}
\tabletypesize{\tiny}
\tablewidth{0pt}
\tablecaption{{\it Chandra} ACIS catalog of the central region of M31}
\tablehead{ID & IAU Name& R.A.& Dec.& Positional &Net&Count Rate&
HR1\,\tablenotemark{a} & HR2\,\tablenotemark{b} & $L_X$\tablenotemark{c} & ~~Notes \\
& (CXOM31) & (h:m:s) & ($^{\circ}$:$'$:$''$) & Error ($''$) &Counts&($10^{-2}$ s$^{-1}$) & & &
($10^{37}$) & }
\startdata
r3-110&J004150.3+411337&00:41:50.336&+41:13:37.16&1.7&91&$0.22\pm0.02$&$0.72\pm0.11$&$-1.0\pm0.01$&0.17&\\
r3-109&J004150.5+412114&00:41:50.514&+41:21:14.66&1.8&93&$0.24\pm0.03$&$0.78\pm0.10$&$0.78\pm0.10$&0.18&\\
r3-81&J004151.5+411438&00:41:51.582&+41:14:38.29&1.1&164&$0.41\pm0.03$&$0.69\pm0.09$&$0.57\pm0.11$&0.31&v\\
r3-108&J004155.0+412303&00:41:55.010&+41:23:03.49&1.3&148&$0.37\pm0.04$&$1.00\pm0.02$&$1.00\pm0.02$&0.28&\\
r3-62&J004204.0+411531&00:42:04.083&+41:15:31.97&0.6&81&$0.20\pm0.02$&$0.68\pm0.13$&$0.27\pm0.24$&0.15&\\
r3-107&J004204.4+410930&00:42:04.435&+41:09:30.63&1.7&75&$0.19\pm0.03$&$0.59\pm0.14$&$0.11\pm0.23$&0.14&\\
r3-80&J004205.7+411133&00:42:05.710&+41:11:33.87&1.1&45&$0.11\pm0.02$&$0.90\pm0.01$&$0.68\pm0.15$&0.08&\\
r3-79&J004207.0+411719&00:42:07.097&+41:17:19.17&0.6&44&$0.11\pm0.02$&$0.74\pm0.15$&$0.70\pm0.17$&0.08&\\
r3-94&J004207.5+411025&00:42:07.500&+41:10:25.63&1.0&71&$0.18\pm0.02$&$0.48\pm0.15$&$0.23\pm0.21$&0.13&v\\
r3-61&J004207.6+411815&00:42:07.697&+41:18:15.17&0.1&1611&$4.06\pm0.10$&$0.64\pm0.04$&$0.51\pm0.05$&3.02&r,v\\
r3-93&J004208.2+411250&00:42:08.278&+41:12:50.92&1.0&32&$0.08\pm0.02$&$1.00\pm0.11$&$1.00\pm0.09$&0.06&\\
r3-60&J004209.0+412048&00:42:09.030&+41:20:48.42&0.3&435&$1.10\pm0.05$&$0.75\pm0.06$&$0.75\pm0.06$&0.82&\\
r3-59&J004209.4+411745&00:42:09.450&+41:17:45.63&0.2&443&$1.12\pm0.05$&$0.71\pm0.06$&$0.56\pm0.08$&0.83&g,r,v\\
r3-102&J004209.6+412008&00:42:09.692&+41:20:08.89&1.0&19&$0.05\pm0.01$&$1.00\pm0.14$&$1.00\pm0.22$&0.04&\\
r3-58&J004210.2+411509&00:42:10.242&+41:15:09.90&0.3&148&$0.37\pm0.03$&$0.58\pm0.10$&$0.28\pm0.16$&0.28&r\\
r3-57&J004210.9+411248&00:42:10.917&+41:12:48.05&0.4&123&$0.31\pm0.03$&$0.11\pm0.15$&$-0.09\pm0.01$&0.23&\\
r3-78&J004211.0+410646&00:42:11.032&+41:06:46.95&1.3&213&$0.54\pm0.04$&$0.26\pm0.12$&$0.39\pm0.12$&0.40&v\\
r3-56&J004211.7+411048&00:42:11.711&+41:10:48.70&0.4&295&$0.74\pm0.04$&$0.64\pm0.07$&$0.59\pm0.09$&0.55&r\\
r3-55&J004211.9+411648&00:42:11.944&+41:16:48.49&0.3&154&$0.39\pm0.03$&$0.86\pm0.06$&$0.80\pm0.08$&0.29&v\\
r3-54&J004212.1+411758&00:42:12.104&+41:17:58.86&0.1&538&$1.36\pm0.06$&$0.72\pm0.06$&$0.67\pm0.06$&1.01&g,r,v\\
r3-92&J004212.7+411244&00:42:12.748&+41:12:44.79&1.0&20&$0.05\pm0.01$&$1.00\pm0.19$&$1.00\pm0.14$&0.04&\\
r3-53&J004213.0+411628&00:42:13.022&+41:16:28.18&0.6&30&$0.07\pm0.01$&$0.66\pm0.20$&$0.15\pm0.47$&0.06&v\\
r3-52&J004213.0+411836&00:42:13.095&+41:18:36.73&0.1&2805&$7.07\pm0.13$&$0.74\pm0.02$&$0.47\pm0.04$&5.27&r,v\\
r3-77&J004214.2+412105&00:42:14.273&+41:21:05.82&0.7&13&$0.03\pm0.01$&$-0.75\pm0.59$&$-0.11\pm0.27$&0.02&v\\
r3-51&J004215.0+412122&00:42:15.029&+41:21:22.04&0.6&15&$0.04\pm0.01$&$1.00\pm0.34$&$1.00\pm0.14$&0.03&\\
r3-50&J004215.0+411234&00:42:15.089&+41:12:34.23&0.1&1017&$2.56\pm0.08$&$0.45\pm0.05$&$0.12\pm0.07$&1.91&r,v,sv\\
r3-49&J004215.2+411802&00:42:15.216&+41:18:02.25&0.5&81&$0.20\pm0.02$&$0.81\pm0.10$&$0.54\pm0.19$&0.15&\\
r3-48&J004215.4+412032&00:42:15.428&+41:20:32.27&0.3&205&$0.52\pm0.04$&$0.64\pm0.08$&$0.38\pm0.14$&0.39&r\\
r3-47&J004215.6+411721&00:42:15.649&+41:17:21.11&0.1&783&$1.97\pm0.07$&$0.57\pm0.06$&$0.28\pm0.08$&1.47&r,v\\
r3-106&J004215.7+412216&00:42:15.772&+41:22:16.18&1.4&21&$0.05\pm0.01$&$0.39\pm0.30$&$-0.32\pm0.06$&0.04&v\\
r3-76&J004216.0+411552&00:42:16.066&+41:15:52.91&0.6&16&$0.04\pm0.01$&$0.54\pm0.40$&$0.88\pm0.57$&0.03&v\\
r3-75&J004216.5+411610&00:42:16.540&+41:16:10.55&0.7&15&$0.04\pm0.01$&$0.82\pm0.17$&$0.80\pm0.49$&0.03&\\
r3-91&J004216.8+411856&00:42:16.886&+41:18:56.89&1.0&17&$0.04\pm0.01$&$0.41\pm0.33$&$0.16\pm0.44$&0.03&\\
r3-46&J004216.9+411508&00:42:16.997&+41:15:08.54&0.3&114&$0.29\pm0.03$&$0.70\pm0.11$&$0.54\pm0.15$&0.21&v,t\\
r3-45&J004218.3+411223&00:42:18.319&+41:12:23.53&0.1&2011&$5.07\pm0.11$&$0.41\pm0.04$&$-0.22\pm0.05$&3.78&r,v\\
r3-44&J004218.6+411401&00:42:18.612&+41:14:01.69&0.1&2428&$6.12\pm0.12$&$0.46\pm0.03$&$0.39\pm0.04$&4.56&g,r,v,sv\\
r3-90&J004218.9+412004&00:42:18.937&+41:20:04.38&0.7&17&$0.04\pm0.01$&$0.83\pm0.15$&$0.54\pm0.38$&0.03&\\
r3-74&J004219.6+412154&00:42:19.657&+41:21:54.07&0.7&38&$0.10\pm0.02$&$0.88\pm0.11$&$0.96\pm0.64$&0.07&g\\
r3-89&J004220.3+411313&00:42:20.313&+41:13:13.98&0.8&15&$0.04\pm0.01$&$0.29\pm0.39$&$-0.34\pm0.07$&0.03&\\
r3-101&J004220.3+410824&00:42:20.368&+41:08:24.78&1.4&34&$0.09\pm0.02$&$0.43\pm0.24$&$0.36\pm0.30$&0.06&\\
r3-43&J004221.0+411808&00:42:21.075&+41:18:08.55&0.3&88&$0.22\pm0.02$&$0.85\pm0.08$&$0.59\pm0.17$&0.17&v,t\\
r3-42&J004221.4+411601&00:42:21.460&+41:16:01.32&0.1&1447&$3.65\pm0.10$&$0.63\pm0.04$&$0.44\pm0.06$&2.72&r,v\\
r3-41&J004221.5+411419&00:42:21.513&+41:14:19.54&0.2&155&$0.39\pm0.03$&$0.28\pm0.13$&$-0.07\pm0.03$&0.29&v\\
r3-40&J004222.3+411333&00:42:22.394&+41:13:33.99&0.1&986&$2.49\pm0.08$&$0.42\pm0.05$&$0.15\pm0.07$&1.85&r,v\\
r3-88&J004222.6+412234&00:42:22.619&+41:22:34.92&0.9&31&$0.08\pm0.02$&$0.35\pm0.26$&$0.30\pm0.30$&0.06&\\
r3-39&J004222.9+411535&00:42:22.919&+41:15:35.14&0.04&2959&$7.46\pm0.14$&$0.51\pm0.03$&$0.37\pm0.04$&5.56&r,v\\
r3-73&J004222.9+410738&00:42:22.958&+41:07:38.21&0.8&129&$0.33\pm0.03$&$0.42\pm0.13$&$0.09\pm0.22$&0.24&\\
r3-38&J004223.1+411407&00:42:23.152&+41:14:07.49&0.1&699&$1.76\pm0.07$&$0.46\pm0.06$&$0.22\pm0.07$&1.31&r,v\\
r2-57&J004224.1+411733&00:42:24.157&+41:17:33.58&0.4&16&$0.04\pm0.01$&$-0.84\pm0.37$&$-0.94\pm0.33$&0.03&\\
r2-52&J004224.1+411535&00:42:24.187&+41:15:35.84&0.3&27&$0.07\pm0.01$&$-0.04\pm0.25$&$-0.08\pm0.17$&0.05&v\\
r2-45&J004225.1+411340&00:42:25.112&+41:13:40.40&0.1&207&$0.52\pm0.04$&$0.22\pm0.12$&$0.00\pm0.13$&0.39&r\\
r2-36&J004226.0+411915&00:42:26.019&+41:19:15.27&0.1&494&$1.25\pm0.06$&$0.50\pm0.07$&$0.26\pm0.09$&0.93&g,r,v\\
r3-87&J004226.1+412552&00:42:26.152&+41:25:52.74&1.0&431&$1.09\pm0.05$&$0.62\pm0.06$&$0.65\pm0.07$&0.81&r\\
r3-37&J004227.6+412048&00:42:27.619&+41:20:48.69&0.7&25&$0.06\pm0.01$&$0.33\pm0.30$&$0.35\pm0.31$&0.05&\\
r3-36&J004228.1+410959&00:42:28.167&+41:09:59.84&0.1&1170&$2.95\pm0.09$&$0.51\pm0.05$&$0.30\pm0.06$&2.20&v\\
r2-35&J004228.2+411222&00:42:28.268&+41:12:22.76&0.04&2865&$7.22\pm0.14$&$0.46\pm0.04$&$0.29\pm0.04$&5.38&r,v,sv\\
r3-111&J004228.8+410434&00:42:28.867&+41:04:34.98&1.0&1783&$4.49\pm0.11$&$0.48\pm0.03$&$0.03\pm0.06$&3.35&r\\
r2-44&J004230.2+411653&00:42:30.244&+41:16:53.41&0.2&15&$0.04\pm0.01$&$0.59\pm0.30$&$0.52\pm0.34$&0.03&\\
r2-43&J004230.9+411910&00:42:30.943&+41:19:10.32&0.2&32&$0.08\pm0.01$&$0.53\pm0.21$&$0.18\pm0.35$&0.06&\\
r2-34&J004231.1+411621&00:42:31.123&+41:16:21.74&0.02&2409&$6.07\pm0.12$&$0.47\pm0.04$&$0.28\pm0.04$&4.52&r,v,sv\\
r2-33&J004231.2+411939&00:42:31.232&+41:19:39.19&0.1&608&$1.53\pm0.06$&$0.55\pm0.06$&$0.34\pm0.08$&1.14&g,r,v\\
r2-51&J004231.2+412008&00:42:31.271&+41:20:08.30&0.3&25&$0.06\pm0.01$&$0.34\pm0.29$&$0.05\pm0.39$&0.05&\\
r3-86&J004231.9+412306&00:42:31.976&+41:23:06.19&1.0&27&$0.07\pm0.02$&$0.76\pm0.19$&$0.62\pm0.25$&0.05&\\
r2-32&J004232.0+411314&00:42:32.057&+41:13:14.24&0.04&1487&$3.75\pm0.28$&$0.05\pm0.03$&$0.50\pm0.05$&2.79&r,sv\\
r2-55&J004232.5+411545&00:42:32.515&+41:15:45.62&0.3&12&$0.03\pm0.01$&$1.00\pm0.44$&$1.00\pm0.21$&0.02&\\
r2-31&J004232.7+411310&00:42:32.726&+41:13:10.77&0.1&306&$0.77\pm0.04$&$0.50\pm0.09$&$0.36\pm0.11$&0.57&v\\
r2-30&J004233.8+411619&00:42:33.884&+41:16:19.88&0.05&362&$0.91\pm0.05$&$0.39\pm0.09$&$0.22\pm0.10$&0.68&r,v\\
r3-35&J004234.1+412150&00:42:34.147&+41:21:50.23&0.3&134&$0.34\pm0.03$&$0.79\pm0.09$&$0.78\pm0.09$&0.25&\\
r2-29&J004234.4+411809&00:42:34.439&+41:18:09.60&0.1&39&$0.10\pm0.02$&$0.28\pm0.25$&$0.13\pm0.29$&0.07&v,t\\
r2-28&J004234.7+411523&00:42:34.771&+41:15:23.32&0.1&165&$0.42\pm0.03$&$0.17\pm0.11$&$-0.45\pm0.15$&0.31&v,t\\
r2-27&J004235.1+412006&00:42:35.199&+41:20:06.09&0.1&385&$0.97\pm0.05$&$0.66\pm0.08$&$0.64\pm0.08$&0.72&r,v\\
r2-42&J004236.5+411350&00:42:36.592&+41:13:50.20&0.2&39&$0.10\pm0.02$&$-0.07\pm0.09$&$-0.81\pm0.22$&0.07&f\\
r3-100&J004237.9+410526&00:42:37.940&+41:05:26.07&1.8&69&$0.17\pm0.03$&$0.65\pm0.14$&$0.00\pm0.61$&0.13&\\
r2-26&J004238.5+411603&00:42:38.581&+41:16:03.80&0.01&7749&$19.53\pm0.22$&$0.57\pm0.02$&$0.67\pm0.02$&14.55&r,v,sv\\
r2-54&J004238.7+411526&00:42:38.770&+41:15:26.44&0.3&12&$0.03\pm0.01$&$0.02\pm0.44$&$-0.65\pm0.48$&0.02&\\
r2-25&J004239.5+411428&00:42:39.529&+41:14:28.52&0.1&287&$0.72\pm0.04$&$0.41\pm0.09$&$-0.17\pm0.13$&0.54&r,v\\
r1-15&J004239.9+411547&00:42:39.986&+41:15:47.68&0.02&1100&$2.77\pm0.08$&$0.32\pm0.05$&$0.03\pm0.06$&2.07&p,r,v,sv\\
r2-24&J004240.1+411845&00:42:40.199&+41:18:45.38&0.1&316&$0.80\pm0.05$&$0.52\pm0.09$&$0.43\pm0.10$&0.59&v\\
r2-23&J004240.5+411355&00:42:40.524&+41:13:55.33&0.1&18&$0.05\pm0.01$&$0.56\pm0.28$&$0.44\pm0.34$&0.03&v\\
r2-22&J004240.6+411327&00:42:40.644&+41:13:27.30&0.04&716&$1.81\pm0.07$&$0.40\pm0.06$&$0.14\pm0.08$&1.34&r,v\\
r3-34&J004240.6+411032&00:42:40.688&+41:10:32.32&0.5&74&$0.19\pm0.02$&$0.36\pm0.18$&$0.37\pm0.18$&0.14&g,v\\
r3-33&J004240.9+412216&00:42:40.912&+41:22:16.58&0.7&22&$0.06\pm0.01$&$-0.37\pm0.35$&$-0.61\pm0.40$&0.04&f\\
r3-85&J004241.0+410701&00:42:41.016&+41:07:01.24&1.1&31&$0.08\pm0.02$&$0.99\pm0.27$&$0.97\pm0.02$&0.06&\\
r3-32&J004241.0+411101&00:42:41.061&+41:11:01.72&0.7&10&$0.02\pm0.01$&$1.00\pm0.26$&$1.00\pm0.44$& 0.02&\\
r1-32&J004241.4+411523&00:42:41.433&+41:15:23.94&0.1&130&$0.33\pm0.03$&$0.42\pm0.12$&$-0.08\pm0.02$& 0.25&g,v\\
r3-31&J004241.6+412106&00:42:41.644&+41:21:06.02&0.3&142&$0.36\pm0.03$&$0.57\pm0.12$&$0.50\pm0.13$& 0.27&v\\
r1-31&J004242.0+411532&00:42:42.076&+41:15:32.18&0.1&90&$0.23\pm0.02$&$0.34\pm0.15$&$-0.32\pm0.25$& 0.17&v\\
r1-5&J004242.1+411608&00:42:42.160&+41:16:08.42&0.01&2779&$7.00\pm0.13$&$0.47\pm0.04$&$0.39\pm0.03$& 5.22&v,t\\
r2-53&J004242.1+411914&00:42:42.171&+41:19:14.20&0.4&8&$0.02\pm0.01$&$0.57\pm0.28$&$0.95\pm0.69$&0.02&\\
r2-21&J004242.3+411445&00:42:42.327&+41:14:45.54&0.05&545&$1.37\pm0.06$&$0.34\pm0.07$&$0.05\pm0.08$&1.02&r,v\\
r1-14&J004242.4+411553&00:42:42.464&+41:15:53.82&0.02&772&$1.95\pm0.07$&$0.45\pm0.06$&$0.29\pm0.07$& 1.45&p,r,v\\
r1-30&J004242.5+411659&00:42:42.515&+41:16:59.59&0.1&70&$0.18\pm0.02$&$0.25\pm0.18$&$0.04\pm0.21$& 0.13&r\\
r2-20&J004242.7+411455&00:42:42.710&+41:14:55.63&0.1&41&$0.10\pm0.02$&$0.03\pm0.25$&$-0.42\pm0.27$&0.08&v\\
r1-13&J004242.9+411543&00:42:42.985&+41:15:43.26&0.03&600&$1.51\pm0.06$&$0.47\pm0.06$&$0.31\pm0.08$& 1.13&r\\
r3-99&J004243.0+410830&00:42:43.050&+41:08:30.48&1.1&24&$0.06\pm0.01$&$0.81\pm0.18$&$0.77\pm0.20$&0.04&\\
r1-24&J004243.1+411640&00:42:43.192&+41:16:40.31&0.1&55&$0.14\pm0.02$&$0.43\pm0.19$&$0.48\pm0.17$& 0.10&p,v\\
r2-19&J004243.3+411319&00:42:43.303&+41:13:19.48&0.1&166&$0.42\pm0.03$&$-0.29\pm0.12$&$-0.96\pm0.07$&0.31&f\\
r1-12&J004243.7+411632&00:42:43.739&+41:16:32.60&0.02&620&$1.56\pm0.06$&$0.41\pm0.07$&$0.04\pm0.08$& 1.16&r,v\\
r1-28&J004243.7+411514&00:42:43.793&+41:15:14.54&0.1&16&$0.04\pm0.01$&$0.31\pm0.38$&$0.27\pm0.35$& 0.03&v\\
r1-29&J004243.7+411629&00:42:43.796&+41:16:29.39&0.1&72&$0.18\pm0.02$&$0.35\pm0.17$&$-0.10\pm0.05$& 0.14&r\\
r1-27&J004243.8+411611&00:42:43.848&+41:16:11.37&0.1&39&$0.10\pm0.02$&$0.36\pm0.23$&$0.21\pm0.25$& 0.07&v\\
r1-23&J004243.8+411604&00:42:43.856&+41:16:04.05&0.04&174&$0.44\pm0.03$&$0.29\pm0.12$&$-0.01\pm0.12$& 0.33&v,t\\
r1-11&J004243.8+411629&00:42:43.883&+41:16:29.97&0.03&273&$0.69\pm0.04$&$0.49\pm0.09$&$0.33\pm0.11$& 0.51&r,v,t\\
r1-33&J004244.0+411604&00:42:44.079&+41:16:04.23&0.1&13&$0.03\pm0.01$&$0.01\pm0.46$&$-0.30\pm0.11$& 0.02&v\\
r1-22&J004244.2+411614&00:42:44.283&+41:16:14.25&0.1&71&$0.18\pm0.02$&$0.57\pm0.14$&$0.13\pm0.18$& 0.13&r\\
r1-10&J004244.3+411608&00:42:44.351&+41:16:08.90&0.03&287&$0.73\pm0.04$&$0.16\pm0.10$&$-0.10\pm0.10$& 0.54&r,v\\
r1-21&J004244.3+411605&00:42:44.355&+41:16:05.56&0.04&197&$0.50\pm0.04$&$0.35\pm0.11$&$-0.06\pm0.02$& 0.37&p,r,v\\
r1-9&J004244.3+411607&00:42:44.365&+41:16:07.65&0.03&286&$0.72\pm0.04$&$0.02\pm0.09$&$-0.85\pm0.08$& 0.54&r,v,t\\
r3-30&J004244.4+411157&00:42:44.404&+41:11:57.90&0.2&209&$0.53\pm0.04$&$0.33\pm0.10$&$0.03\pm0.14$& 0.39&v\\
r1-8&J004244.6+411618&00:42:44.667&+41:16:18.28&0.04&234&$0.59\pm0.04$&$0.14\pm0.11$&$-0.34\pm0.12$& 0.44&r\\
r3-29&J004244.8+411137&00:42:44.844&+41:11:37.76&0.1&1306&$3.29\pm0.09$&$0.33\pm0.04$&$-0.12\pm0.06$& 2.45&r,v\\
r2-18&J004244.8+411739&00:42:44.895&+41:17:39.82&0.1&185&$0.47\pm0.03$&$0.47\pm0.11$&$0.24\pm0.14$&0.35&\\
r1-26&J004245.0+411523&00:42:45.073&+41:15:23.30&0.1&111&$0.28\pm0.03$&$0.21\pm0.15$&$-0.09\pm0.01$& 0.21&p\\
r2-17&J004245.0+411407&00:42:45.092&+41:14:07.13&0.1&106&$0.27\pm0.03$&$0.30\pm0.15$&$0.08\pm0.09$&0.18&r\\
r1-4&J004245.1+411621&00:42:45.109&+41:16:21.90&0.02&908&$2.29\pm0.08$&$0.33\pm0.06$&$0.13\pm0.07$& 1.70&r,v\\
r2-16&J004245.2+411722&00:42:45.219&+41:17:22.49&0.1&308&$0.78\pm0.04$&$0.41\pm0.09$&$0.29\pm0.10$&0.58&v,t\\
r1-20&J004245.2+411611&00:42:45.226&+41:16:11.31&0.1&125&$0.32\pm0.03$&$0.09\pm0.14$&$-0.29\pm0.17$& 0.24&v\\
r1-7&J004245.5+411608&00:42:45.586&+41:16:08.79&0.04&164&$0.41\pm0.03$&$0.38\pm0.11$&$0.12\pm0.13$& 0.31&v\\
r3-72&J004245.7+412434&00:42:45.751&+41:24:34.46&0.8&50&$0.13\pm0.02$&$0.21\pm0.24$&$0.38\pm0.22$&0.09&\\
r1-19&J004245.9+411619&00:42:45.996&+41:16:19.77&0.1&105&$0.26\pm0.03$&$0.29\pm0.15$&$0.00\pm0.16$& 0.20&v,t\\
r2-15&J004246.0+411736&00:42:46.067&+41:17:36.35&0.1&48&$0.12\pm0.02$&$0.39\pm0.22$&$0.35\pm0.23$&0.09&g,v\\
r1-18&J004246.1+411543&00:42:46.144&+41:15:43.37&0.1&88&$0.22\pm0.02$&$0.46\pm0.14$&$0.05\pm0.20$& 0.17&r\\
r3-28&J004246.9+412119&00:42:46.912&+41:21:19.65&0.3&79&$0.20\pm0.02$&$0.77\pm0.10$&$0.74\pm0.12$& 0.15&\\
r1-3&J004246.9+411615&00:42:46.963&+41:16:15.76&0.02&934&$2.35\pm0.08$&$0.44\pm0.05$&$0.10\pm0.07$& 1.75&r,v\\
r1-2&J004247.1+411628&00:42:47.167&+41:16:28.65&0.01&2600&$6.55\pm0.13$&$0.50\pm0.04$&$0.38\pm0.04$& 4.88&p,r,v,sv\\
r3-27&J004247.2+411157&00:42:47.232&+41:11:57.75&0.3&84&$0.21\pm0.02$&$0.40\pm0.16$&$0.09\pm0.21$& 0.16&v\\
r3-26&J004247.8+411052&00:42:47.863&+41:10:52.42&0.5&23&$0.06\pm0.01$&$0.64\pm0.24$&$0.44\pm0.32$& 0.04&\\
r1-17&J004247.8+411623&00:42:47.865&+41:16:23.19&0.05&138&$0.35\pm0.03$&$0.50\pm0.12$&$0.21\pm0.17$& 0.26&v\\
r1-25&J004247.8+411549&00:42:47.888&+41:15:49.85&0.1&26&$0.07\pm0.01$&$-0.79\pm0.22$&$-1.00\pm0.01$& 0.05&\\
r1-6&J004247.8+411533&00:42:47.891&+41:15:33.04&0.02&1002&$2.53\pm0.08$&$0.22\pm0.05$&$-0.14\pm0.06$& 1.88&r\\
r3-25&J004248.5+412523&00:42:48.501&+41:25:23.10&0.3&574&$1.45\pm0.06$&$0.66\pm0.05$&$0.51\pm0.07$& 1.08&r,v\\
r1-1&J004248.5+411521&00:42:48.528&+41:15:21.31&0.01&3360&$8.47\pm0.15$&$0.45\pm0.03$&$0.32\pm0.03$& 6.31&r,v,sv\\
r1-16&J004248.7+411624&00:42:48.715&+41:16:24.64&0.05&144&$0.36\pm0.03$&$0.26\pm0.13$&$-0.07\pm0.02$& 0.27&v\\
r3-84&J004248.9+412406&00:42:48.970&+41:24:06.87&0.8&59&$0.15\pm0.02$&$-0.86\pm0.14$&$-1.00\pm0.01$&0.11&v\\
r2-41&J004249.1+411742&00:42:49.141&+41:17:42.25&0.2&15&$0.04\pm0.01$&$-0.05\pm0.31$&$-0.11\pm0.24$&0.03&\\
r2-14&J004249.2+411816&00:42:49.233&+41:18:16.26&0.1&322&$0.81\pm0.05$&$0.50\pm0.08$&$0.39\pm0.11$&0.60&r,v\\
r3-98&J004249.3+410635&00:42:49.389&+41:06:35.69&1.1&37&$0.09\pm0.02$&$0.66\pm0.19$&$-0.06\pm0.51$&0.07&\\
r3-24&J004249.9+411108&00:42:49.991&+41:11:08.64&0.5&38&$0.09\pm0.02$&$0.29\pm0.35$&$0.82\pm0.13$& 0.07&\\
r2-40&J004250.2+411813&00:42:50.228&+41:18:13.11&0.3&11&$0.03\pm0.01$&$0.72\pm0.28$&$0.78\pm0.18$&0.02&\\
r2-56&J004250.4+411556&00:42:50.476&+41:15:56.15&0.3&12&$0.03\pm0.01$&$-0.72\pm0.41$&$-1.0\pm0.01$&0.02&p\\
r3-71&J004250.7+411033&00:42:50.732&+41:10:33.58&0.7&28&$0.07\pm0.01$&$-0.26\pm0.18$&$-0.09\pm0.15$&0.05&g\\
r3-70&J004251.5+412633&00:42:51.539&+41:26:33.58&1.0&80&$0.69\pm0.08$&$0.34\pm0.18$&$0.47\pm0.17$&0.51&\\
r2-39&J004251.6+411302&00:42:51.633&+41:13:02.65&0.2&46&$0.12\pm0.02$&$0.57\pm0.18$&$0.33\pm0.27$&0.09&\\
r2-50&J004252.2+411735&00:42:52.281&+41:17:35.02&0.2&22&$0.05\pm0.01$&$0.52\pm0.26$&$0.07\pm0.42$&0.04&\\
r2-49&J004252.5+411835&00:42:52.526&+41:18:35.24&0.3&12&$0.03\pm0.01$&$1.00\pm0.43$&$1.00\pm0.22$&0.02&\\
r2-12&J004252.5+411540&00:42:52.528&+41:15:40.20&0.03&918&$2.31\pm0.08$&$-0.98\pm0.01$&$-1.0\pm0.01$&1.72&r,v\\
r2-13&J004252.5+411854&00:42:52.528&+41:18:54.75&0.02&3943&$9.94\pm0.16$&$0.52\pm0.03$&$0.39\pm0.04$&7.40&r,v\\
r2-38&J004252.6+411328&00:42:52.621&+41:13:28.72&0.2&13&$0.03\pm0.01$&$0.18\pm0.42$&$0.09\pm0.45$&0.02&\\
r3-69&J004253.6+412553&00:42:53.656&+41:25:53.56&0.7&188&$0.47\pm0.04$&$-0.38\pm0.11$&$-0.97\pm0.10$&0.35&r,s\\
r2-11&J004254.9+411603&00:42:54.937&+41:16:03.46&0.02&3282&$8.27\pm0.14$&$0.44\pm0.03$&$0.30\pm0.03$&6.16&r,v,sv\\
r2-10&J004255.1+411836&00:42:55.186&+41:18:36.33&0.1&179&$0.45\pm0.03$&$0.42\pm0.11$&$-0.07\pm0.02$&0.34&r,v\\
r3-23&J004255.3+412556&00:42:55.394&+41:25:56.60&0.3&1011&$2.55\pm0.08$&$0.64\pm0.04$&$0.52\pm0.06$& 1.90&r,v\\
r2-9&J004255.6+411835&00:42:55.628&+41:18:35.44&0.1&108&$0.27\pm0.03$&$0.26\pm0.15$&$-0.08\pm0.02$&0.20&g,r,v\\
r2-8&J004256.9+411844&00:42:56.914&+41:18:44.37&0.1&286&$0.72\pm0.04$&$0.42\pm0.09$&$0.17\pm0.12$&0.54&v,t\\
r3-22&J004257.9+411104&00:42:57.932&+41:11:04.59&0.1&1646&$4.15\pm0.10$&$0.43\pm0.04$&$0.12\pm0.06$& 3.09&r\\
r2-7&J004258.3+411529&00:42:58.335&+41:15:29.46&0.1&487&$1.23\pm0.06$&$0.42\pm0.07$&$0.00\pm0.10$&0.91&r,v\\
r2-59&J004258.6+411200&00:42:58.621&+41:12:00.14&0.6&15&$0.04\pm0.01$&$0.03\pm0.80$&$0.90\pm0.29$&0.03&\\
r2-58&J004259.0+411158&00:42:59.093&+41:11:58.75&0.6&14&$0.03\pm0.01$&$1.00\pm0.39$&$1.00\pm0.18$&0.03&\\
r3-97&J004259.1+412613&00:42:59.160&+41:26:13.76&1.2&130&$0.33\pm0.03$&$0.51\pm0.11$&$0.16\pm0.20$&0.24&\\
r2-48&J004259.5+411242&00:42:59.558&+41:12:42.26&0.3&41&$0.10\pm0.02$&$0.34\pm0.25$&$0.34\pm0.25$&0.08&\\
r2-6&J004259.6+411919&00:42:59.672&+41:19:19.72&0.04&1678&$4.23\pm0.10$&$0.42\pm0.04$&$0.17\pm0.05$&3.15&g,r,v,sv\\
r2-5&J004259.8+411606&00:42:59.881&+41:16:06.01&0.03&1356&$3.42\pm0.09$&$0.46\pm0.05$&$0.39\pm0.05$&2.55&g,r,v\\
r2-37&J004301.1+411351&00:43:01.119&+41:13:51.69&0.2&65&$0.16\pm0.02$&$0.48\pm0.17$&$0.24\pm0.23$&0.12&r\\
r2-47&J004301.7+411814&00:43:01.716&+41:18:14.98&0.2&15&$0.04\pm0.01$&$0.25\pm0.38$&$-0.37\pm0.17$&0.03&v\\
r3-96&J004301.7+411052&00:43:01.754&+41:10:52.63&0.8&15&$0.04\pm0.01$&$-0.34\pm0.23$&$-0.63\pm0.52$&0.03&\\
r2-46&J004301.8+411726&00:43:01.821&+41:17:26.78&0.3&12&$0.03\pm0.01$&$-0.35\pm0.27$&$-1.00\pm0.01$&0.02&f\\
r3-68&J004302.4+411203&00:43:02.421&+41:12:03.71&0.9&17&$0.04\pm0.01$&$1.00\pm0.35$&$1.00\pm0.15$&0.03&\\
r2-4&J004302.9+411522&00:43:02.955&+41:15:22.82&0.04&1430&$3.60\pm0.10$&$0.43\pm0.05$&$0.23\pm0.06$&2.69&g,r,v\\
r3-21&J004303.0+412042&00:43:03.036&+41:20:42.54&0.3&84&$0.21\pm0.02$&$0.78\pm0.10$&$0.65\pm0.15$& 0.16&p\\
r3-20&J004303.1+411015&00:43:03.167&+41:10:15.18&0.2&284&$0.72\pm0.04$&$0.24\pm0.10$&$0.14\pm0.12$& 0.53&r\\
r2-3&J004303.2+411528&00:43:03.241&+41:15:28.00&0.04&1249&$3.15\pm0.09$&$0.34\pm0.05$&$-0.03\pm0.01$&2.35&r,v,t\\
r3-19&J004303.3+412122&00:43:03.309&+41:21:22.42&0.1&632&$1.59\pm0.06$&$0.36\pm0.06$&$-0.05\pm0.02$& 1.19&g,r,v\\
r2-2&J004303.8+411805&00:43:03.890&+41:18:05.23&0.05&1174&$2.96\pm0.09$&$0.45\pm0.05$&$0.16\pm0.06$&2.21&g,r,v,sv\\
r2-1&J004304.2+411601&00:43:04.264&+41:16:01.62&0.1&249&$0.63\pm0.04$&$0.33\pm0.10$&$-0.05\pm0.03$&0.47&v\\
r3-83&J004306.6+412243&00:43:06.696&+41:22:43.82&0.9&27&$0.07\pm0.02$&$0.05\pm0.35$&$0.26\pm0.03$&0.05&e\\
r3-67&J004306.8+411912&00:43:06.801&+41:19:12.26&0.6&32&$0.08\pm0.02$&$-0.06\pm0.17$&$-0.28\pm0.23$&0.06&\\
r3-18&J004307.5+412020&00:43:07.550&+41:20:20.09&0.2&195&$0.49\pm0.04$&$0.38\pm0.11$&$0.08\pm0.15$& 0.37&g,v\\
r3-95&J004307.8+412418&00:43:07.816&+41:24:18.16&1.2&33&$0.08\pm0.02$&$0.04\pm0.27$&$0.19\pm0.32$&0.06&\\
r3-17&J004308.6+411248&00:43:08.687&+41:12:48.30&0.2&320&$0.81\pm0.05$&$0.55\pm0.09$&$0.60\pm0.08$& 0.60&r\\
r3-16&J004309.8+411901&00:43:09.869&+41:19:01.22&0.2&396&$1.00\pm0.05$&$0.47\pm0.08$&$0.22\pm0.10$& 0.74&r,v,t\\
r3-15&J004310.6+411451&00:43:10.665&+41:14:51.55&0.04&4432&$11.17\pm0.17$&$0.41\pm0.03$&$0.29\pm0.03$&8.32&g,r,v\\
r3-14&J004311.3+411809&00:43:11.391&+41:18:09.76&0.5&43&$0.11\pm0.02$&$0.98\pm0.50$&$0.99\pm0.77$& 0.08&\\
r3-13&J004313.2+411813&00:43:13.250&+41:18:13.96&0.3&103&$0.26\pm0.03$&$0.33\pm0.15$&$0.02\pm0.20$& 0.19&v\\
r3-12&J004313.9+411712&00:43:13.986&+41:17:12.19&0.8&22&$0.05\pm0.01$&$0.33\pm0.34$&$0.36\pm0.34$& 0.04&\\
r3-112&J004314.3+410725&00:43:14.323&+41:07:25.42&0.8&1051&$2.65\pm0.08$&$0.86\pm0.04$&$0.94\pm0.02$&1.97&g,r\\
r3-11&J004314.3+411651&00:43:14.344&+41:16:51.31&0.6&42&$0.11\pm0.02$&$0.36\pm0.22$&$-0.25\pm0.13$& 0.08&\\
r3-105&J004314.5+412513&00:43:14.585&+41:25:13.66&1.5&62&$0.16\pm0.02$&$0.36\pm0.17$&$0.04\pm0.22$&0.12&g\\
r3-10&J004315.4+411125&00:43:15.499&+41:11:25.00&0.6&77&$0.19\pm0.02$&$0.35\pm0.16$&$-0.06\pm0.15$& 0.14&g,r\\
r3-9&J004316.1+411841&00:43:16.165&+41:18:41.73&0.3&166&$0.42\pm0.03$&$0.51\pm0.11$&$0.29\pm0.15$& 0.31&v\\
r3-8&J004318.8+412018&00:43:18.851&+41:20:18.52&0.6&85& $0.21\pm0.03$&$-0.96\pm0.10$&$-1.00\pm0.01$& 0.16&v\\
r3-66&J004320.9+411852&00:43:20.922&+41:18:52.62&0.7&18&$0.04\pm0.01$&$0.23\pm0.38$&$-0.49\pm0.34$&0.03&\\
r3-7&J004321.1+411751&00:43:21.136&+41:17:51.19&0.4&194&$0.49\pm0.04$&$0.41\pm0.10$&$0.06\pm0.15$& 0.37&p,r,v\\
r3-104&J004321.5+411558&00:43:21.510&+41:15:58.34&1.0&19&$0.05\pm0.01$&$0.68\pm0.28$&$0.38\pm0.34$&0.04&\\
r3-82&J004322.2+411258&00:43:22.283&+41:12:58.67&1.0&16&$0.04\pm0.01$&$0.90\pm0.13$&$0.87\pm0.19$&0.03&\\
r3-65&J004324.1+411314&00:43:24.148&+41:13:14.34&0.9&28&$0.07\pm0.02$&$1.00\pm0.15$&$1.00\pm0.10$&0.05&\\
r3-6&J004324.9+411728&00:43:24.957&+41:17:28.33&0.5&101&$0.25\pm0.03$&$0.71\pm0.10$&$0.71\pm0.10$& 0.19&\\
r3-5&J004326.0+411935&00:43:26.026&+41:19:35.88&0.4& 34&$0.08\pm0.02$&$-0.33\pm0.29$&$-0.73\pm0.40$& 0.06\\
r3-4&J004326.3+411756&00:43:26.328&+41:17:56.87&0.8&22&$0.05\pm0.01$&$0.07\pm0.30$&$-0.64\pm0.60$& 0.04&\\
r3-64&J004326.3+411910&00:43:26.376&+41:19:10.82&1.0&82&$0.21\pm0.03$&$0.67\pm0.13$&$0.83\pm0.09$&0.15&\\
r3-63&J004327.8+411829&00:43:27.871&+41:18:29.83&0.6&258&$0.65\pm0.04$&$-0.46\pm0.09$&$-1.00\pm0.01$&0.48&r,s\\
r3-103&J004329.1+410749&00:43:29.116&+41:07:49.11&2.6&1387&$3.49\pm0.10$&$0.81\pm0.03$&$0.85\pm0.03$&2.60&r\\
r3-3&J004332.4+411041&00:43:32.460&+41:10:41.66&0.4&1419&$3.58\pm0.10$&$0.86\pm0.03$&$0.94\pm0.16$& 2.66&\\
r3-2&J004334.4+411323&00:43:34.410&+41:13:23.72&0.4&994&$2.51\pm0.08$&$0.38\pm0.05$&$-0.04\pm0.01$& 1.87&r,v\\
r3-1&J004337.2+411443&00:43:37.269&+41:14:43.07&0.4&2264&$5.71\pm0.12$&$0.56\pm0.03$&$0.49\pm0.04$& 4.25&g,v\\
\enddata

\normalsize
\tablenotetext{a}{HR1 = (M--S)/(M+S)}
\tablenotetext{b}{HR2 = (H--S)/(H+S)}
\tablenotetext{c}{Luminosity in 0.3--7 keV (erg s$^{-1}$), assuiming
an absorbed power-law model with a photon index of 1.7 and $N_H=10^{21}$ cm$^{-2}$.}
\tablecomments{e: Extragalactic objects; f: Foreground stars; g: GC; p:
PN; r: {\it ROSAT} HRI sources; s: SNR; v: Variables; sv: Spectral
variables; t:transients} 

\end{deluxetable}

%% file: table4.tex
\begin{deluxetable}{lcllc}
\tablewidth{0pt}
\tablecaption{Optical IDs}
\tablehead{ID & {\it Chandra}& Type & Identification & Radial \\
& Name&&& Offset ($''$)}
\startdata

r3-59&CXOM31\,J004209.4+411745&GC&mita140& 1.0\\
r3-54&CXOM31\,J004212.1+411758&GC&Bo78,mita153 &2.1,0.9\\
r3-44&CXOM31\,J004218.6+411401&GC& Bo86,mit164&1.1,0.7\\
r3-74&CXOM31\,J004219.6+412154&GC&mita165,166&0.8,0.7\\
r2-36&CXOM31\,J004226.0+411915&GC&Bo96,mita174 &1.6,1.2\\
r2-33&CXOM31\,J004231.2+411939&GC& Bo107,mita192&0.8,0.9\\
r2-42&CXOM31\,J004236.5+411350 & Star & Ha94(238126) & 0.1\\
r1-15&CXOM31\,J004239.9+411547&PN&Ford17&2.9\\
r3-34&CXOM31\,J004240.6+411032&GC&Bo123,mita212&2.9,0.8\\
r3-33&CXOM31\,J004240.9+412216 & Star & Ha94(278717) & 0.6\\
r1-32&CXOM31\,J004241.4+411523&GC&mita213&1.5\\
r1-14&CXOM31\,J004242.4+411553&PN& Ford4&2.9\\
r1-24&CXOM31\,J004243.1+411640&PN&Ford322&0.6\\
r2-19&CXOM31\,J004243.3+411319& Star & Ha94(235849) & 0.5\\
r1-21&CXOM31\,J004244.3+411605&PN&Ford316&2.9\\
r1-33&CXOM31\,J004244.0+411604& &&1.6\\
r1-26&CXOM31\,J004245.0+411523&PN&Ford21&0.8\\
r2-15&CXOM31\,J004246.0+411736&GC&PB-in7&0.7\\
r1-2 &CXOM31\,J004247.1+411628 & PN & Ford13 & 1.0\\
r2-56&CXOM31\,J004250.4+411556 & PN & Ford462 & 0.5\\ 
r3-71&CXOM31\,J004250.7+411033&GC&mita222,PB-in2&0.5,1.2\\
r3-69&CXOM31\,J004253.6+412553&SNR&DO80(13)&6.0\\
r2-9&CXOM31\,J004255.6+411835&GC&Bo138&2.5\\
r2-6&CXOM31\,J004259.6+411919&GC&Bo143&1.7\\
r2-5&CXOM31\,J004259.8+411606&GC&Bo144&2.2\\
r2-46&CXOM31\,J004301.8+411726 & Star & Ha94(261262) & 0.6\\
r2-4&CXOM31\,J004302.9+411522&GC&Bo146&1.9\\
r3-21&CXOM31\,J004303.0+412042 & PN & Ford201 & 2.4\\
r3-19&CXOM31\,J004303.3+412122&GC&Bo147,mita240&2.5,1.3\\
r2-2&CXOM31\,J004303.8+411805 &GC&Bo148&2.5\\
r3-83&CXOM31\,J004306.6+412243&EO&MLA93(686)&0.9\\
r3-18&CXOM31\,J004307.5+412020&GC&Bo150,mita246&2.8,1.1\\
r3-15&CXOM31\,J004310.6+411451&GC&Bo153,mita251&1.9,0.4\\
r3-112&CXOM31\,J004314.3+410725&GC&Bo158&2.3\\
r3-105&CXOM31\,J004314.5+412513&GC&Bo159,mita258&1.4,0.9\\
r3-10&CXOM31\,J004315.4+411125&GC&mita260&0.9\\
r3-7&CXOM31\,J004321.1+411751 & PN & Ford209 & 1.7\\
r3-63&CXOM31\,J004327.8+411829&SNR&DO80(15)&5.1\\
r3-1&CXOM31\,J004337.2+411443&GC&mita299&0.6\\
\enddata

\tablecomments{Bo--Battistini et al. (1987); mita--Magnier (1993);
Ford--Ford78, Ci89; PB--Barmby (2001); MLA--SIMBAD} 
\end{deluxetable}

%% file: table5.tex
\begin{deluxetable}{lccc}
\tablewidth{0pt}
\tablecaption{Cross-correlation of {\it Chandra} and {\it ROSAT} sources}
\tablehead{ID & {\it Chandra} Name& {\it ROSAT} HRI& Radial
\\
& & Identification& Offset ($''$)}
\startdata
r3-61&CXOM31\,J004207.6+411815& PF93(4) & 0.5\\
r3-59&CXOM31\,J004209.4+411745& PF93(5) &2.0 \\
r3-58&CXOM31\,J004210.2+411509& PF93(6) & 1.6\\
r3-56&CXOM31\,J004211.7+411048& PF93(7) & 1.0\\
r3-54&CXOM31\,J004212.1+411758& PF93(8) & 2.1\\
r3-52&CXOM31\,J004213.0+411836& PF93(9) & 1.5\\
r3-50&CXOM31\,J004215.0+411234& PF93(10) & 2.9\\
r3-48&CXOM31\,J004215.4+412032& PF93(11) & 0.8\\
r3-47&CXOM31\,J004215.6+411721& PF93(12) & 1.5\\
r3-45&CXOM31\,J004218.3+411223& PF93(14) & 1.5\\
r3-44&CXOM31\,J004218.6+411401& PF93(15) & 1.2\\
r3-42&CXOM31\,J004221.4+411601& PF93(17) & 1.5\\
r3-40&CXOM31\,J004222.3+411333& PF93(18) & 1.5\\
r3-39&CXOM31\,J004222.9+411535& PF93(20) & 1.5\\
r3-38&CXOM31\,J004223.1+411407& PF93(21) & 1.7\\
r2-45&CXOM31\,J004225.1+411340& PF93(22) & 1.2\\
r2-36&CXOM31\,J004226.0+411915& PF93(23) & 1.6\\
r3-87&CXOM31\,J004226.1+412552& PF93(24) & 0.9\\
r2-35&CXOM31\,J004228.2+411222& PF93(25) & 2.4\\
r3-111&CXOM31\,J004228.8+410434& PF93(26) & 3.5\\
r2-34&CXOM31\,J004231.1+411621& PF93(27) & 1.5\\
r2-33&CXOM31\,J004231.2+411939& PF93(28) & 0.8\\
r2-32&CXOM31\,J004232.0+411314& PF93(29) & 1.6\\
r2-30&CXOM31\,J004233.8+411619& PF93(32) & 5.5\\
r2-27&CXOM31\,J004235.1+412006& PF93(34) & 1.0\\
r2-26&CXOM31\,J004238.5+411603& PF93(35) & 1.5\\
r2-25&CXOM31\,J004239.5+411428& PF93(36) & 1.8\\
r1-15&CXOM31\,J004239.9+411547& PF93(37) & 1.7\\
r2-22&CXOM31\,J004240.6+411327& PF93(38) & 1.5\\
r2-21&CXOM31\,J004242.3+411445& PF93(39) & 0.8\\
r1-14&CXOM31\,J004242.4+411553& PF93(41) & 2.1\\
r1-30&CXOM31\,J004242.5+411659& PF93(40) & 1.4\\
r1-13&CXOM31\,J004242.9+411543& PF93(42) & 2.0\\
r1-12&CXOM31\,J004243.7+411632& PF93(43) & 0.5\\
r1-11&CXOM31\,J004243.8+411629& & 3.4\\
r1-29&CXOM31\,J004243.7+411629&& 3.7\\
r1-10&CXOM31\,J004244.3+411608& PF93(44) & 1.0\\
r1-9 &CXOM31\,J004244.3+411607 & &1.9\\
r1-21&CXOM31\,J004244.3+411605 & &3.9\\
r1-22&CXOM31\,J004244.2+411614 & &5.1\\
r1-8&CXOM31\,J004244.6+411618& PF93(47) & 1.2\\
r1-4&CXOM31\,J004245.1+411621 & &5.1\\
r3-29&CXOM31\,J004244.8+411137& PF93(45) & 1.7\\
r2-17&CXOM31\,J004245.0+411407& PF93(46) & 1.9\\
r1-3&CXOM31\,J004246.9+411615&  & 2.7\\
r1-18&CXOM31\,J004246.1+411543& PF93(48) & 4.1\\
r1-2&CXOM31\,J004247.1+411628& PF93(50) & 1.7\\
r1-6&CXOM31\,J004247.8+411533& PF93(52) & 1.8\\
r3-25&CXOM31\,J004248.5+412523& PF93(53) & 2.6\\
r1-1&CXOM31\,J004248.5+411521& PF93(54) & 1.9\\
r2-14&CXOM31\,J004249.2+411816& PF93(55) & 1.0\\
r2-12&CXOM31\,J004252.5+411540& PF93(58) & 1.3\\
r2-13&CXOM31\,J004252.5+411854& PF93(57) & 1.2\\
r3-69&CXOM31\,J004253.6+412553& PF93(59) & 1.8\\
r2-11&CXOM31\,J004254.9+411603& PF93(60) & 1.8\\
r3-23&CXOM31\,J004255.3+412556& PF93(61) & 0.2\\
r2-9&CXOM31\,J004255.6+411835& PF93(62) & 2.5\\
r2-10&CXOM31\,J004255.1+411836 & & 4.3\\
r3-22&CXOM31\,J004257.9+411104& PF93(64) & 3.2\\
r2-7&CXOM31\,J004258.3+411529& PF93(65) & 1.6\\
r2-6&CXOM31\,J004259.6+411919& PF93(66) & 1.6\\
r2-5&CXOM31\,J004259.8+411606& PF93(67) & 2.2\\
r2-37&CXOM31\,J004301.1+411351& PF93(68) & 2.7\\
r2-4&CXOM31\,J004302.9+411522& PF93(70) & 1.9\\
r2-3&CXOM31\,J004303.2+411528 &&4.3\\
r3-20&CXOM31\,J004303.1+411015& PF93(71) & 1.5\\
r3-19&CXOM31\,J004303.3+412122& PF93(72) & 2.5\\
r2-2&CXOM31\,J004303.8+411805& PF93(73) & 2.5\\
r3-17&CXOM31\,J004308.6+411248& PF93(74) & 3.7\\
r3-16&CXOM31\,J004309.8+411901& PF93(75) & 2.1\\
r3-15&CXOM31\,J004310.6+411451& PF93(76) & 1.9\\
r3-112&CXOM31\,J004314.3+410725& PF93(77) & 2.3\\
r3-10&CXOM31\,J004315.4+411125& PF93(78) & 3.9\\
r3-7&CXOM31\,J004321.1+411751& PF93(79) & 2.3\\
r3-63&CXOM31\,J004327.8+411829& PF93(80) & 1.7\\
r3-103&CXOM31\,J004329.1+410749& PF93(81) & 2.3\\
r3-2&CXOM31\,J004334.4+411323& PF93(82) & 5.6\\
\enddata
\end{deluxetable}